\documentstyle[11pt]{article}

\makeatletter
\@addtoreset{equation}{section}
\makeatother
\renewcommand{\theequation}{\thesection.\arabic{equation}}

\def\dalemb#1#2{{\vbox{\hrule height .#2pt
        \hbox{\vrule width.#2pt height#1pt \kern#1pt
                \vrule width.#2pt}
        \hrule height.#2pt}}}

\def\td{\tilde}

 \def\bd{\begin{document}} \def\ed{\end{document}}
\def\ds{\documentstyle} \let\fr=\frac \let\bl=\bigl \let\br=\bigr
\let\Br=\Bigr \let\Bl=\Bigl
\let\bm=\bibitem
\let\na=\nabla
\let\pa=\partial \let\ov=\overline
\newcommand{\be}{\begin{equation}}
\newcommand{\ee}{\end{equation}}
\def\ba{\begin{array}}
\def\ea{\end{array}}
\def\ft#1#2{{\textstyle{{\scriptstyle #1}\over {\scriptstyle #2}}}}
\def\fft#1#2{{#1 \over #2}}
\def\del{\partial}
\def\sst#1{{\scriptscriptstyle #1}}
\def\oneone{\rlap 1\mkern4mu{\rm l}}
\newcommand{\ho}[1]{$\, ^{#1}$}
\newcommand{\hoch}[1]{$\, ^{#1}$}
\newcommand{\bea}{\begin{eqnarray}}
\newcommand{\eea}{\end{eqnarray}}
\newcommand{\ra}{\rightarrow}
\newcommand{\lra}{\longrightarrow}
\newcommand{\Lra}{\Leftrightarrow}
\newcommand{\ap}{\alpha^\prime}
\newcommand{\bp}{\tilde \beta^\prime}
\newcommand{\tr}{{\rm tr} }
\newcommand{\Tr}{{\rm Tr} }
\newcommand{\NP}{Nucl. Phys. }
\newcommand{\tamphys}{\it Center for Theoretical Physics,
Texas A\&M University, College Station, Texas 77843}

\newcommand{\auth}{H. L\"u and C.N. Pope}

\begin{document}

\hfill{CTP-TAMU-47/95}

\hfill{hep-th/9512012}

\vspace{20pt}

\begin{center}
{ \large {\bf $p$-brane Solitons in Maximal Supergravities}}

\vspace{30pt}
\auth

\vspace{15pt}
{\tamphys}
\vspace{40pt}

\underline{ABSTRACT}
\end{center}
\vspace{40pt}

    In this paper, we give a construction of $p$-brane solitons in all
maximal supergravity theories in $4\le D \le 11$ dimensions that are
obtainable from $D=11$ supergravity by dimensional reduction.  We first
obtain the full bosonic Lagrangians for all these theories in a formalism
adapted to the $p$-brane soliton construction.  The solutions that we
consider involve one dilaton field and one antisymmetric tensor field
strength, which are in general linear combinations of the basic fields of
the supergravity theories.   We also study the supersymmetry properties of
the solutions by calculating the eigenvalues of the Bogomol'nyi matrices,
which are derived from the commutators of the supercharges.   We give an
exhaustive list of the supersymmetric $p$-brane solutions using field
strengths of all degrees $n=4,3,2,1$, and the non-supersymmetric solutions
for $n=4,3,2$.  As well as studying elementary and solitonic solutions, we
also discuss dyonic solutions in $D=6$ and $D=4$.  In particular, we find
that the Bogomol'nyi matrices for the supersymmetric massless dyonic
solutions have indefinite signature.

{\vfill\leftline{}\vfill
\vskip	10pt
\footnoterule
{\footnotesize	Research supported in part by DOE
Grant DE-FG05-91-ER40633 \vskip	-12pt}}

\pagebreak
\setcounter{page}{1}

\section{Introduction}

     Recent progress in the understanding of duality and the
non-perturbative structure of string theories has emphasised the importance
of solitonic string and $p$-brane solutions in the low-energy effective
actions.  It is therefore of interest to attempt to classify all such
solutions.  Our goal in this paper is to study $p$-brane solitons in all the
maximal supergravities related to 11-dimensional supergravity or,
equivalently, type IIA supergravity in 10 dimensions, by dimensional
reduction.   Starting from the 11-dimensional theory, whose bosonic sector
comprises the vielbein and a 4-index antisymmetric tensor field strength,
one obtains theories by dimensional reduction containing a vielbein,
dilatonic scalar fields, and various $n$-index antisymmetric tensor field
strengths $F_n$ with $n=1,2,3,4$.  The $p$-brane solutions we shall be
considering are of the kind discussed in [1-9], and involve non-vanishing
background values for the vielbein, a dilatonic scalar, and an $n$-index
field strength.  The dilatonic scalar and the field strength, which we shall
refer to as the canonical dilaton and field strength, may be linear
combinations of the original fields in the supergravity theory.   The
relevant part of the $D$-dimensional Lagrangian is given by
\be
{\cal L} = e
R -\ft12 e\, (\del\phi)^2 -\fft1{2n!} e\, e^{a\phi} F_n^2 \ .\label{genlag}
\ee

     We shall be focusing on isotropic $p$-brane solutions, for which the
metric ansatz is given by
\be
ds^2 = e^{2A}\, dx^\mu dx^\nu \eta_{\mu\nu} + e^{2B}\, dy^mdy^m
\ ,\label{metricform}
\ee
where $x^\mu\, (\mu=0, \ldots, d-1)$ are the coordinates of the
$(d-1)$-brane world volume, and $y^m$ are the coordinates of the
$(D-d)$-dimensional tranverse space.  The functions $A$ and $B$, as well as
the dilaton $\phi$, depend only on $r=\sqrt{y^m y^m}$.  Thus the ansatz
preserves an $SO(1, d-1)\times SO(D-d)$ subgroup of the original $SO(1,
D-1)$ Lorentz group.   The constant $a$ in the dilaton prefactor can be
parametrised as \cite{lpss}
\be
a^2 = \Delta - \fft{2d\tilde d} {D-2}\ ,\label{avalue}
\ee
where $\tilde d\equiv D-d -2$ and $d\tilde d = (n-1)(D-n-1)$.   In $D=11$,
the absence of a dilaton implies that $\Delta = 4$.  In fact this value occurs
in the dilaton prefactors for all field strengths obtained simply by
Kaluza-Klein dimensional reduction, since the reduction procedure always
preserves the value of $\Delta$ \cite{lpss}.  However, if the field strength
$F_n$ used in a particular $p$-brane solution is formed from a linear
combination of these original field strengths, then it will have a dilaton
prefactor which, after setting the non-participating fields to zero, has
$\Delta < 4$.

      For each $F_n$ with $n\ge 2$, there are two different ans\"atze that
also preserve the same subgroup, namely \cite{dabl,pew2}
\be
F_{r\mu_1\ldots\mu_{n-1}} = \epsilon_{\mu_1\ldots\mu_{n-1}} (e^C)'\qquad
{\rm or}\qquad
F_{m_1\ldots m_n} = \lambda \epsilon_{m_1\ldots m_np}\,
\fft{y^p}{r^{n+1}}\ ,\label{fansatz}
\ee
where a prime denotes a derivative with respect to $r$.  The first case
gives rise to an elementary $(d-1)$-brane with $d=n-1$, and the second
gives rise to a solitonic $(d-1)$-brane with $d= D-n-1$.  The solutions are
given by \cite{lpss}
\bea
ds^2 &=& \Big (1+\fft{k}{r^{\tilde d}}\Big)^{-\ft{4\tilde d}{\Delta(D-2)}}
\, dx^\mu dx^\nu \eta_{\mu\nu} + \Big(1+ \fft{k}{r^{\tilde d}}\Big)^{
\ft{4d}{\Delta(D-2)}}\, dy^m dy^m\ ,\nonumber\\
e^\phi &=& \Big (1+\fft{k}{r^{\tilde d}}\Big)^{
\ft{2a}{\epsilon \Delta}}\ , \label{gensol}
\eea
where $\epsilon = 1$ and $-1$ for the elementary and the
solitonic solutions respectively, and $k= - \sqrt{\Delta}
\lambda/(2 \tilde d)$.  In the elementary case, the function $C$ satisfies
the equation
\be
e^C = \fft{2}{\sqrt\Delta} \Big ( 1+ \fft{k}{r^{\tilde d}}\Big)^{-1}\ .
\ee
Note that the dual of the solution for the field strength in the elementary
case is identical to the field strength of the solitonic case, and {\it vice
versa}.  For this reason, we shall only consider solutions for field
strengths with $n\le D/2$.

     So far we have discussed the solutions (\ref{gensol}) for an $n$-form
field strength with $n > 1$. When $n=1$, and hence $\tilde d=0$, one can
only construct a solitonic solution, given by eqn.\ (\ref{gensol}) with
$kr^{-\tilde d} \longrightarrow k\log{r}$ and $\tilde d \longrightarrow 0$.
The solution (\ref{gensol}) is also valid when $n=0$, giving rise to a
purely dilatonic $(D-2)$-brane in $D$ dimensions \cite{pw,lpss2}, which we
shall not consider in this paper.

     For a given value of $n$, the metrics of the elementary and solitonic
$p$-brane solutions (\ref{gensol}) are characterised completely by the value
of $\Delta$.  However, as we shall see later, inequivalent field-strength
configurations can sometimes give rise to the same value of $\Delta$.  These
solutions, although having the same value of $\Delta$, break different
fractions of the $D=11$ supersymmetry.   The purpose of this paper is to
classify the possible $p$-brane solutions that can arise from a maximal
supergravity theory in any dimension $D$ according to their values of
$\Delta$ and their field-strength configurations.   We shall also examine
the supersymmetry of these solutions.

      In dimensions $D=2n$, the field strength $F_n$ can in principle
use both the elementary and solitonic ans\"atze (\ref{fansatz})
simultaneously.  In this case, the equations of motion reduce to two
independent differential equations
\be
\phi'' - n \fft{\phi'}{r} =\ft12 a (S_1^2 - S_2^2)\ ,\qquad
A'' + n \fft{A'}{r} = \ft14 (S_1^2 + S_2^2)\ ,\label{phiadiff}
\ee
together with the relations $B=-A$, $(e^C)' = \lambda_1 e^{a\phi + 2(n-1)A}
r^{-n}$, where $S_1$ and $S_2$ are given by
\be
S_1 = \lambda_1 e^{\ft12 a \phi + (n-1)A} r^{-n}\ ,\qquad
S_2 = \lambda_2 e^{-\ft12 a\phi + (n-1)A} r^{-n}\ .\label{s1s2}
\ee
The equations (\ref{phiadiff}) admit a simple solution either when
$a^2=n-1$, given by
\be
e^{-\ft12 a \phi -(n-1) A} = 1 + \ft{\lambda_1}{a\sqrt2} r^{-n+1}\ ,\qquad
e^{+\ft12 a \phi -(n-1) A} = 1 + \ft{\lambda_2}{a\sqrt2} r^{-n+1}\ ,
\label{dyonicsol1}
\ee
or when $a=0$, given by
\be
\phi = 0\ ,\qquad e^{- (n-1) A} = 1 +\ft12
\sqrt{\ft{\lambda_1^2+\lambda_2^2}{(n-1)}} r^{-n+1}\ .\label{dyonicsol2}
\ee
The former solution includes the dyonic strings in $D=6$ that were found in
ref.\ \cite{dfkr}, and the latter solution includes the self-dual string in
$D=6$ \cite{dkl}.  The dyonic solutions (\ref{dyonicsol1}) and
(\ref{dyonicsol2}) apply only to $a^2=n-1$ and $a^2=0$, and hence $\Delta =
2n-2$ and $\Delta=n-1$, respectively.  In $D=8, 6$ and 4 dimensions, the
values of $\Delta$ are $\{6,3\}$, $\{4,2\}$ and $\{2,1\}$ respectively.
Thus such a dyonic
solution does not exist in maximal supergravity in $D=8$ since the 4-form
has $\Delta=4$.  We shall see in the next section that such a dyonic
solution is also excluded by consideration of the full set of field
equations of the $D=8$ supergravity theory.  Note that the solution
(\ref{dyonicsol1}) becomes the standard elementary or solitonic solution
(\ref{gensol}) when $\lambda_2=0$ or $\lambda_1=0$ respectively.  When
$\lambda_1=\lambda_2$, the solutions (\ref{dyonicsol1}) and
(\ref{dyonicsol2}) are equivalent.

     The paper is organised as follows.  In section 2, we obtain the
complete bosonic Lagrangian of the maximal supergravities in lower
dimensions {\it via} Kaluza-Klein dimensional reduction from $D=11$.  We do
this in a formalism that distinguishes between the dilatonic scalar fields,
which appear in exponential prefactors, and the 0-form potentials for 1-form
field strengths.   In section 3, we first discuss the formalism for
constructing $p$-brane solutions in the maximal supergravity theories.
The supersymmetry properties of these solutions can be examined by
constructing the appropriate Nester forms, which arise as the commutators of
conserved supercharges.  We obtain the Nester forms in lower dimensions
from the Nester form in $D=11$, by using the Kaluza-Klein procedure.  We can
read off the ``Bogomol'nyi matrices'' from these Nester forms.  Zero
eigenvalues of the the Bogomol'nyi matrix correspond to unbroken
supersymmetries in the corresponding $p$-brane solutions.   In sections 4, 5
and 6, we obtain the explicit $p$-brane solutions corresponding to the
4-form, 3-form, 2-form and 1-form field strengths, and we discuss their
supersymmetry properties.   We present the conclusions in section 7.
Details of the dimensionally-reduced bosonic Lagrangians can be found in an
appendix.

\section{Kaluza-Klein dimensional reduction}

     A convenient way of constructing the maximal supergravity theories in
arbitrary dimensions is by starting from $D=11$ supergravity and performing
Kaluza-Klein dimensional reduction.  Although one can directly reduce from
11 to $D$ dimensions, there are advantages in carrying out the procedure
step by step, descending through the dimensions one at a time.  In this
approach, it is easier to identify which of the scalar fields are
``dilatonic,'' {\it i.e.\ }scalar fields that appear in the Lagrangian {\it
via} exponential factors, as opposed to those that have constant shift
symmetries, which should be viewed as 0-form potentials for 1-form field
strengths.  Thus the dimensional reduction procedure consists of an
iterative application of the basic one-step reduction from $(D+1)$ to $D$
dimensions.   We denote the coordinates of the $(D+1)$-dimensional spacetime
by $x^{\hat {\sst M}} = (x^{\sst M}, z)$, where $z$ is the coordinate of the
extra dimension.  The vielbein of
the $(D+1)$-dimensional spacetime is then given by $\hat
e^{\sst A}=e^{\alpha\varphi}\, e^{\sst A}$ and $\hat e^{\underline z}=
e^{-(D-2)\alpha
\varphi} \, (dz +{\cal A})$, where $e^{\sst A}$ is the vielbein in $(D+1)$
dimensions.  In terms of components, $\hat e^{\hat{\sst A}}{}_{\hat{\sst M}}$
is given by
\bea
\hat e^{\sst A}{}_{\sst M} = e^{\alpha \varphi}\, e^{\sst A}{}_{\sst M}\ ,
&& \hat e^{\underline z}{}_{\sst M} = e^{-(D-2)\alpha \varphi} {\cal
A}_{\sst M}\ ,\nonumber\\
\hat e^{\sst A}{}_z =0\ ,&& \hat e^{\underline z}{}_z = e^{-(D-2)
\alpha\varphi}\ ,\label{vbd+1}
\eea
where $e^{\sst A}{}_{\sst M}$, $\varphi$ and ${\cal A}={\cal A}_{\sst M}
dx^{\sst M}$ are taken to be independent of the extra coordinate $z$, and
${\sst M}$ and $z$ denote world indices whilst ${\sst A}$ and ${\underline
z}$ denote tangent-space indices.   The constant $\alpha$ is given by
$\alpha^2 = \fft1{2(D-1)(D-2)}$.   An $n$-index field strength $\hat F_n$ in
$(D+1)$ decomposes into two $z$-independent field strengths $F_n$ and
$F_{n-1}$ in $D$ dimensions:
\be
\hat F_n = F_n + F_{n-1} \wedge (dz + {\cal A}) \ .
\label{transgress}
\ee
Note that the $D$-dimensional field strengths are in general all of the form
$F = d A + \cdots$, where the dots indicate extra ``Chern-Simons'' terms
involving wedge products of lower degree forms and potentials.   Under
certain circumstances, these modifications to the field strengths will lead
to constraints on the allowable field configurations for $p$-brane
solutions; we shall discuss this in more detail later.  Under this reduction
procedure, a bosonic Lagrangian in $(D+1)$ dimensions of the form
\be
{\cal L} = \hat e\hat R -\ft12\hat e\, (\del \phi)^2 -\fft1{2n!}\hat e\,
e^{\hat a \hat\phi} \,\hat F_n^2\label{d+1lag}
\ee
becomes
\bea
{\cal L} &=& e R -\ft12 e\, (\del\phi)^2 -\ft12 e\, (\del\varphi)^2
-\ft14e\, e^{-2(D-1)\alpha \phi} {\cal F}^2\nonumber\\
&& -\fft1{2n!}e\, e^{-2(n-1)\alpha \varphi +\hat a\phi} F_n^2 -
\fft1{2(n-1)!} e\, e^{2(D-n)\alpha \varphi +\hat a \phi} F_{n-1}^2\ ,
\label{dlag}
\eea
where ${\cal F}= d {\cal A}$.  One can apply this procedure iteratively to
reduce the 11-dimensional supergravity to any lower dimension.  Note that
in each step of dimensional reduction, a new two-form field strength emerges
from the metric.

    As one descends through the dimensions, the number of field strengths
proliferates.  The dimensional reduction of the Lagrangian (\ref{d+1lag}) to
(\ref{dlag}) described above indicates the pattern of this proliferation. In
each step, a new dilatonic scalar appears, and so there will be $(11-D)$
dilatons in $D$ dimensions.   Thus in $(D+1)$ dimensions, the dilaton factor
of the field strength $F_n$ in Lagrangian (\ref{d+1lag}) in general takes
the form $e^{\vec a_{\sst{D+1}} \cdot \vec \phi_{\sst{D+1}}}\, \hat F_n^2$,
where $\vec \phi_{\sst{D+1}} = (\phi_1, \phi_2, \ldots, \phi_{\sst{10-D}})
$.  Since there is no dilaton in 11-dimensional supergravity, we see that
$\vec a_{11} =0$.   It follows from eqn.\ (\ref{dlag}) that we obtain the
lower dimensional vectors $\vec a$ by the following algorithm:
\be
\vec a_{\sst D} = \Big(\vec a_{\sst{D+1}},\, x\,
\sqrt{\ft2{(D-1)(D-2)}}
\,\Big)\quad {\rm with}\quad
x= \cases{
-(n-1),  & for $\hat F_n \rightarrow F_n$, \cr
(D-n), & for $\hat F_n \rightarrow F_{n-1}$,\cr
-(D-1), & for ${\cal F}$.\cr}\label{algorithm}
\ee
Here the three cases refer to an $n$-form field strength coming from an
$n$-form in $(D+1)$ dimensions, an $(n-1)$-form coming from an $n$-form, and
the 2-form coming from the metric.  In this last case, since ${\cal F}$
appears for the first time, the vector $\vec a_{\sst{D+1}}$ is zero.  At the
next step of the reduction, this 2-form behaves just like any other 2-form.
This algorithm gives a complete construction of all the dilaton prefactors
in any maximal $D$-dimensional supergravity, starting from 11-dimensional
supergravity with its single 4-form field strength.   For example, in
$D=10$, there is one 4-form, one 3-form and one 2-form, and their (one
component) dilaton vectors are given by $(-\ft12)$, $(1)$ and $(-\ft32)$
respectively.   In $D=9$, there is one 4-form whose vector is $(-\ft12,
-\ft3{2\sqrt7})$; two 3-forms with $(-\ft12, \ft5{2\sqrt7})$ and $(1,
-\ft1{\sqrt7})$; three 2-forms with $(1, \ft3{\sqrt7})$, $(-\ft32,
-\ft1{2\sqrt7})$ and $ (0, -\ft4{\sqrt7})$; and one 1-form with
$(-\ft32, \ft7{2\sqrt7})$.

      In general, we find that the solution to the recursion relation
(\ref{algorithm}) is as follows.   Let us denote the dilaton vectors for the
4-form $F_{\sst{MNPQ}}$, the 3-forms $F_{\sst{MNP}i}$, the 2-forms
$F_{\sst{MN}ij}$ and the 1-forms $F_{\sst{M}ijk}$ by $\vec a$, $\vec a_{i}$,
$\vec a_{ij}$ and $\vec a_{ijk}$ respectively, where $i$ labels the internal
$(11-D)$ indices in $D$ dimensions.  The index runs from $i=1$,
corresponding to the dimension that is compactified in going from $D=11$ to
$D=10$, down to $i = (11-D)$.  There are also 2-forms ${\cal
F}_{\sst{MN}}^{(i)}$ and 1-forms ${\cal F}_{\sst{M}i}^{(j)}$ with $i<j$,
coming from the dimensional reduction of the vielbein.  We denote their
dilaton vectors by $\vec b_i$ and $\vec b_{ij}$ respectively.  We find that
the dilaton vectors are given by
\bea
&&F_{\sst{MNPQ}}\qquad\qquad\qquad\qquad\qquad\qquad
{\rm vielbein}\nonumber\\
{\rm 4-form:}&&\vec a = -\vec g\ ,\nonumber\\
{\rm 3-forms:}&&\vec a_i = \vec f_i -\vec g \ ,\label{dilatonvec}\\
{\rm 2-forms:}&& \vec a_{ij} = \vec f_i + \vec f_j - \vec g\ ,
\qquad\qquad\qquad \,\,\, \,\vec b_i = -\vec f_i\nonumber\,\\
{\rm 1-forms:}&&\vec a_{ijk} = \vec f_i + \vec f_j + \vec f_k -\vec g
\ ,\qquad\qquad\vec b_{ij} = -\vec f_i + \vec f_j\ ,\nonumber
\eea
where the vectors $\vec g$ and $\vec f_i$ have $(11-D)$ components
in $D$ dimensions, and are given by
\bea
\vec g &=&3 (s_1, s_2, \ldots, s_{11-\sst D})\ ,\nonumber\\
\vec f_i &=& \Big(\underbrace{0,0,\ldots, 0}_{i-1}, (10-i) s_i, s_{i+1},
s_{i+2}, \ldots, s_{11-\sst D}\Big)\ ,
\eea
where $s_i = \sqrt{2/((10-i)(9-i))}$.  It is easy to see that they satisfy
\be
\vec g \cdot \vec g = \ft{2(11-D)}{D-2}, \qquad
\vec g \cdot \vec f_i = \ft{6}{D-2}\ ,\qquad
\vec f_i \cdot \vec f_j = 2\delta_{ij} + \ft2{D-2}\ .\label{gfdot}
\ee
Note that the definitions in (\ref{dilatonvec}) are given for $i<j<k$, and
that the vectors $\vec a_{ij}$ and $\vec a_{ijk}$ are antisymmetric in their
indices. The 1-forms ${\cal F}_{\sst{M}i}^{(j)}$ and hence the vectors
$b_{ij}$ are only defined for $i<j$, but it is sometimes convenient to
regard them as being antisymmetric too, by defining $\vec b_{ij}=-\vec b_{ji}$
for $i>j$.  Eqn.\ (\ref{dilatonvec}), together
with (\ref{gfdot}), contains all the necessary information about the dilaton
vectors in $D$-dimensional maximal supergravity.

    The bosonic Lagrangian of $D=11$ supergravity is \cite{cj}
\be
{\cal L} = \hat e \hat R -\ft1{48} \hat e\, \hat F_4^2 +\ft16 \hat F_4\wedge
\hat F_4\wedge \hat A_3
\ .\label{d11lag}
\ee
(We are representing the final term as an 11-form rather than a 0-form in
order to avoid writing out the $\epsilon$ tensor and all the associated
indices.  It is understood that the last term should be dualised.)  The
bosonic Lagrangian for maximal supergravity in $D$ dimensions, obtained by
dimensional reduction of (\ref{d11lag}), is therefore
\bea
{\cal L} &=& eR -\ft12 e\, (\del\vec\phi)^2 -\ft1{48}e\, e^{\vec a\cdot \vec
\phi}\, F_4^2 -\ft{1}{12} e\sum_i e^{\vec a_i\cdot \vec\phi}\, (F_3^{i})^2
-\ft14 e\, \sum_{i<j} e^{\vec a_{ij}\cdot \vec\phi}\, (F_2^{ij})^2
\nonumber\\
&& -\ft14e\, \sum_i e^{\vec b_i\cdot \vec\phi}\, ({\cal F}_2^i)^2
-\ft12 e\, \sum_{i<j<k} e^{\vec a_{ijk} \cdot\vec \phi}\,
(F_1^{ijk})^2 -\ft12e\, \sum_{i<j} e^{\vec b_{ij}\cdot \vec\phi}\,
({\cal F}_1^{ij})^2 + {\cal L}_{\sst{FFA}}\ ,\label{dgenlag}
\eea
where $F_4$, $F_3^i$, $F_2^{ij}$ and $F_1^{ijk}$ are the 4-form, 3-forms,
2-forms and 1-forms coming from the dimensional reduction of $\hat F_4$ in
$D=11$; ${\cal F}_2^i$ are the 2-forms coming from the dimensional reduction
of the vielbein, and ${\cal F}_1^{ij}$ are the 1-forms coming from the
dimensional reduction of these 2-forms.  In general the field strengths
appearing in the kinetic terms acquire Chern-Simons type modifications in
the dimensional reduction process.  In the appendix, we give the complete
expressions for these modified field strengths.  We denote the modified
fields by untilded symbols, and the unmodifed fields, $\tilde F_4 = d A_3$,
{\it etc.}, by tilded symbols.  The final term ${\cal L}_{\sst{FFA}}$ in
(\ref{dgenlag}) comes from the dimensional reduction of $\hat F_4\wedge \hat
F_4\wedge \hat A_3$, and is also given in the appendix.  Note that in
some lower dimensions, certain higher-degree forms can be dualised to
lower-degree forms. We shall always do this, so that in $D$ dimensions all
forms have degree $n\le D/2$. The dilaton vectors of these dualised forms
are equal to the negatives of the original dilaton vectors \cite{dkl}.

    Our approach to finding $p$-brane solutions in the maximal supergravity
theories is first to solve the equations following from (\ref{genlag}), and
then to select only those solutions that in addition satisfy the constraints
implied by both ${\cal L}_{\sst{FFA}}$ and the Chern-Simons modifications of
the field strengths of the kinetic terms.  First we discuss the constraints
from ${\cal L}_{\sst{FFA}}$. These arise in cases where the 0-form
potentials $A_0^{ijk}$ appear linearly in ${\cal L}_{\sst{FFA}}$ in a
particular dimension.  In order to set the $A_0^{ijk}$ potentials to zero
consistently with their equations of motion, it is therefore necessary that the
bilinear product of field strengths that occurs  multiplied by $A_0^{ijk}$ in
${\cal L}_{\sst{FFA}}$ should vanish.  This imposes the following
constraints:
\bea
D=8: &&  F_4\wedge F_4= 0\ ,\nonumber\\
D=7: &&  F_4 \wedge F_3^i = 0\ ,\nonumber\\
D=6: &&  F_4 \wedge F_2^{ij} = 0\ ,\qquad
F^i_3\wedge F^j_3= 0\ ,\label{ffaconstr}\\
D=5: && F_3^{[i}\wedge F_2^{jk]} = 0\ ,\nonumber\\
D=4: && F_2^{[ij}\wedge F_2^{kl]} = 0\ .\nonumber
\eea
If one considers $p$-brane solutions with a purely elementary or a purely
solitonic ansatz on the original fields, then the above constraints are
identically satisfied.  However, in $D\le 8$ dimensions, the higher-degree
field strengths can be dualised to lower-degree field strengths (or else
they dualise to field strengths of the same degree).  Thus the elementary
(or solitonic) ansatz for a dualised field strength is equivalent to the
solitonic (or elementary) ansatz for the original field strength.  When the
participating field strengths for a $p$-brane solution mix dualised and
undualised fields, the above constraints can become non-trivial.   For
example, in $D=7$ we cannot have a $p$-brane solution simultaneously
involving both the 3-forms $F_3^i$ and the 3-form $*F_4$ coming from the
dualisation of the 4-form.  Similarly we cannot have $*F_4$ and $F_2^{ij}$
non-zero at the same time in $D=6$.  On the other hand, the constraints
involving both 2-forms $F_2^{ij}$ and the dualised 2-forms $*F_3^i$ from
3-forms in $D=5$ are satisified as long as  each $F_2^{ij}$ has an index in
common with each $*F_3^k$. Similar considerations apply to the 0-brane
solutions in $D=4$, where, as we may see from (\ref{ffaconstr}), if some of
the 2-forms $F_2^{ij}$ have elementary contributions and others have
solitonic contributions, then each of the former type must have an index in
common with each of the latter type.

      Since the elementary and solitonic ans\"atze (\ref{fansatz}) both have
the property that the field strength is closed, $d F=0$, it follows that in
our $p$-brane solutions of the supergravity theories, we must require that
all the Chern-Simons modifications to the field strengths should preserve
this property.  In addition, field strengths of degrees other than the
degree of the fields participating in the solution must be zero.  It
therefore follows that the exterior derivatives of all the Chern-Simons
modifications must vanish.  This gives rise to non-trivial constraints in
$D=5$ and $D=4$. For example, consider $0$-brane (particle) solutions using
2-forms in $D=4$. Since the forms of all other degrees are assumed to be
zero, this means in particular that the 0-form potentials ${\cal A}^{ij}_0$
are zero, and hence $\gamma^{ij}=\delta^{ij}$ in (\ref{A.6}). It follows
that the vanishing of the 3-forms $F^i_3$ gives rise to the non-trivial
constraint
\be
F_2^{ij}\wedge {\cal F}_2^{j} = 0\ .\label{csconstr}
\ee
For the purely elementary or purely solitonic ansatz, this constraint is
identically satisfied.  However, when the solution has mixed elementary and
solitonic contributions, the constraint vanishes only when the corresponding
field strengths do not have a common index.

      In this paper, our principal interest is in the $p$-brane solutions
with either a pure elementary or a pure solitonic ansatz for each field
strength. In certain dimensions, where a field strength dualises to one of
the same degree, we can also in principle consider $p$-brane solutions in
which the canonical field strength has both elementary and solitonic
contributions. These dyonic configurations may or may not satisfy the
constraints implied by the ${\cal L}_{\sst{FFA}}$ term and the
Chern-Simons modifications to the field strengths. For example, we can do
this for the 3-form field strengths in $D=6$ and for the 2-form field
strengths in $D=4$. The situation is different in $D=8$, where such a mixed
elementary and solitonic solution does not exist for the 4-form field
strength, as can be easily seen from the above constraint. (However, the
situation is different if a non-zero $A_0^{ijk}$ is allowed; in fact such a
solution does exist \cite{ilpt}, but it lies outside the class of solutions
that we are considering in this paper. In all the solutions we are
considering, there is no non-zero contribution from ${\cal L}_{\sst{FFA}}$.)
 Finally, we remark that when one considers $p$-brane solutions for 1-form
field strengths in $D=5$, the equations of motion for the 0-form potentials
$A_0^{ijk}$ are modified by an $F_4\wedge F_1^{ijk}$ term.  This additional
term vanishes if we only consider $p$-brane solutions that do not involve
the dualised 1-form $*F_4$.

\section{General solutions and Bogomol'nyi matrices}

      We have seen in the introduction that the metric of a $p$-brane
solution is specified once the $\Delta$ of the dilaton prefactor of the
field strength is given.  We have stated also that the values of $\Delta$
for all the original field strengths that are obtained simply by
Kaluza-Klein dimensional reduction are given by $\Delta =4$.  (It is very
easy to verify this by substituting $\vec a\cdot \vec a$ into eqn.\
(\ref{avalue}), using the expression for the $\vec a$'s that we obtained in
the previous section.) Thus if we use any one of these field strengths to
construct a $p$-brane solution while setting all the other field strengths
to zero, it will have $\Delta =4$. However, as we shall now show, there are
other solutions in which a linear combination of field strengths of a given
degree, and a corresponding linear combination of the dilatons, are
involved.  These solutions will in general have $\Delta < 4$.   The
possibility of making such linear combinations depends crucially on the
properties of the dot products of the dilaton vectors for the various field
strengths.  Let us consider a bosonic Lagrangian with $N$ $n$-forms
$F_\alpha$, with $\alpha =1, 2, \ldots N$, namely
\be
{\cal L} = eR -\ft12 e\, (\del \vec\phi)^2 -\fft1{2n!}
\sum_{\alpha =1}^{N} e^{\vec a_{\alpha}\cdot \vec\phi}\, F_\alpha^2
\ .\label{nlag}
\ee
To find a solution involving all these field strengths and a single linear
combination $\phi$ of dilaton fields, we perform a rotation in the space of
dilatons by writing
\be
\vec \phi = \vec n\, \phi + \vec \phi_{\perp}
\quad {\rm such\  that}\quad \vec a_\alpha \cdot \vec n = a\
\quad {\rm for\  all}\quad \alpha\ ,
\label{phirot}
\ee
where $\vec n$ is a unit vector and $\vec n\cdot \vec\phi_{\perp}=0$.  Note
that the requirement that all the $\vec a_\alpha$ have the same projection
on $\vec n$ is necessary so that the participating field strengths have a
Lagrangian of the form (\ref{genlag}), with a common exponential prefactor.
In order to be able to set $\vec \phi_{\perp}$ to zero consistently with its
equations of motion, it is therefore necessary that
\be
\sum_{\alpha} \vec a_{\alpha} F_{\alpha}^2 = a \vec n \sum_\alpha
F^2_\alpha \ .\label{fequation1}
\ee
Taking the dot product with $\vec a_\beta$, we obtain
\be
\sum_\alpha A_{\alpha\beta} F_\alpha^2 = a^2 \sum_\alpha F_\alpha^2\ ,
\label{fequation2}
\ee
where $A_{\alpha\beta}\equiv \vec a_\alpha \cdot \vec a_\beta$.  If the
matrix $A_{\alpha\beta}$ is invertible, we therefore have
\bea
F_\beta^2 &=& a^2 \sum_\alpha (A^{-1})_{\alpha\beta} \sum_\gamma F_\gamma^2
\ ,\nonumber\\
a^2 &=& \Big( \sum_{\alpha, \beta} (A^{-1})_{\alpha\beta} \Big)^{-1}
\ ,\label{avaluesol}
\eea
which gives rise to a $p$-brane solution with $\Delta = \Big( \sum_{\alpha,
\beta} (A^{-1})_{\alpha\beta} \Big)^{-1} + 2d\tilde d/(D-2)$.  In this
case, we can easily see that $a \ne 0$, and hence the unit vector $\vec n$
can be found from eqn (\ref{fequation1}).

    When the matrix $A_{\alpha\beta}$ is singular,  the analysis is
different.  Unlike the the previous case, we can have solutions to the
homogeneous equation (\ref{fequation2}) with more than just the one
rescaling parameter. We shall first consider solutions with only the
rescaling parameter.  In this case, the solutions must have $a=0$ since the
matrix $A_{\alpha\beta}$ is singular and has a zero eigenvector, which
solves eqn.\ (\ref{fequation2}).  This corresponds to $\Delta = 2{d\tilde
d}/(D-2)$. When a solution to eqn.\ (\ref{fequation2}) has extra free
parameters in addition to the overall scale, we can have values of $a$ other
than zero.  However, the values that occur are still discrete, and
independent of the values of the parameters. Thus these values of $a$ are
just repetitions of those that occurred for cases with fewer than $N$
participating field strengths, since we may without loss of generality
choose the parameters such that one or more of the field strengths vanishes,
thereby reducing the number of participating fields to a previously
considered case.   For the purpose of this paper, therefore, we need only
consider the cases where $A_{\alpha\beta}$ is non-singular, or, if it is
singular, it should have only one zero eigenvalue, leading to $a=0$.  A
particular class of singular $A_{\alpha\beta}$'s are the ones corresponding
to the cases when $N$ is larger than the number of dilatonic scalar fields.
In this case, we find that all the solutions involve more than the one
overall scale parameter, and thus need not be considered further.  Thus in
dimension $D$, we need consider at most $(11-D)$ non-vanishing field
strengths.  Note that equations (\ref{fequation1}) and (\ref{fequation2})
determine the relations between the squares of the field strengths, rather
than the field strengths themselves.  The remaining freedom to choose the
signs of the corresponding Page charges sometimes affects the supersymmetry
properties of the solutions, as we shall see later.

      In the next three sections, we shall carry out the search for solutions
in all dimensions $D\ge 4$, for all $n$-forms with $n\le 4$.  For each
solution, we shall also examine its supersymmetry.  In order to determine
how much supersymmetry is preserved by the solutions, we now
develop a general formalism using the Bogomol'nyi matrix, {\it i.e.}\ the
commutator of conserved supercharges.   For each component of unbroken
supersymmetry, this matrix will have a zero eigenvalue.  We can construct
the Bogomol'nyi matrices by first constructing the one for 11-dimensional
supergravity, and then obtaining those for $D<11$ by dimensional reduction.

     Given a spinor $\epsilon$ that is asymptotically constant as
$r\rightarrow \infty$, the associated supercharge per unit spatial
$p$-volume of the the $p$-brane is given by \cite{dabl}
\be
Q_\epsilon = \int_{\del\Sigma} \bar\epsilon \Gamma^{\sst{ABC}}
\psi_{\sst C}\, d\Sigma_{\sst{AB}}\ ,\label{supercharge}
\ee
where $\del\Sigma$ is the $(\tilde d+1)$-sphere with radius $r$ in the
transverse space.  The commutator of the resulting supercharges is given by
\be
[Q_{\epsilon_1}, \, Q_{\epsilon_2}] = \delta_{\epsilon_1}\, Q_{\epsilon_2}
=\int_{\del \Sigma} N^{\sst{AB}}\, d\Sigma_{\sst{AB}}\ ,
\label{commutator}
\ee
where $N^{\sst{AB}} = \bar \epsilon_1\Gamma^{\sst{ABC}}\delta_{\epsilon_2}
\psi_{\sst C}$.   From the transformation rule for the gravitino in $D=11$
supergravity, we therefore obtain the Nester form
\be
N^{\sst{AB}} = \bar \epsilon_1 \Gamma^{\sst{ABC}}\, D_{\sst C} \epsilon_2 +
\ft18 \bar \epsilon_1 \Gamma^{\sst{C_1C_2}}\epsilon_2\,\,
F^{\sst{AB}}{}_{\sst{C_1C_2}} + \ft1{96} \bar \epsilon_1
\Gamma^{\sst{ABC_1\ldots C_4}}\epsilon_2\,\, F_{\sst{C_1\ldots C_4}}\ .
\label{nestor}
\ee
Since only the $d\Sigma_{0r}$ component of the $p$-brane spatial volume
element contributes, we may read off the Bogomol'nyi matrix ${\cal M}$ from
the integral
\be
 {1\over \omega_{\tilde d+1}}   \int_{\del\Sigma\, {\rm at}\,
r\rightarrow \infty} N^{0r} r^{\tilde d +1} d\Omega_{(\tilde d+1)}
 = \epsilon_1^\dagger {\cal M} \epsilon_2\ ,
\label{matdef}
\ee
where $\omega_{\tilde d + 1}$ is the volume of the unit $(\tilde d +1)$-sphere.
If there is an unbroken supersymmetry, then there exists a Killing spinor
such that eqn.\ (\ref{commutator}) vanishes.  In other words, the
Bogomol'nyi matrix (\ref{matdef}) has a zero eigenvalue for each component
of the unbroken supersymmetry.

      We can now use the Bogomol'nyi matrix to study the supersymmetry of
the $p$-brane solutions in $D=11$ dimensions.  There is only one field
strength in $D=11$ supergravity, namely the 4-form, which gives rise to an
elementary membrane and a solitonic 5-brane whose metrics are given by
(\ref{gensol}).  Note that for the elementary ansatz for field strengths
given by (\ref{fansatz}), the last term in (\ref{nestor}) vanishes, whilst
the second term vanishes for the solitonic ansatz.  Substituting the
solutions into eqn.\ (\ref{matdef}), we obtain
\bea
{\rm elementary:} && {\cal M} = m\oneone + u \Gamma_{012}\ ,\nonumber\\
{\rm solitonic:} && {\cal M} = m\oneone + v \Gamma_{{\hat 1}{\hat 2}
{\hat 3}{\hat 4}{\hat 5}}\ ,\label{bog11}
\eea
where the hats indicate index values in the transverse space, while indices
without hats live in the world-brane volume.  The parameter $m$ denotes the
mass per unit volume of the $p$-brane, and $u=\ft{1}{4\omega_{7}} \int_{S^7}
*F$ and $v=\ft{1}{4\omega_{4}}\int_{S^4} F$ are precisely the electric and
magnetic Page charges of the field strength \cite{page}. They are given by
\be
m = \fft{\lambda}{4}\ ,\qquad u =\fft{\lambda}{4} = v\ .
\label{masscharge}
\ee
The eigenvalues of the Bogomol'nyi matrix (\ref{bog11}) can be easily obtained
without needing to decompose the 11-dimensional $\Gamma$ matrices, by
invoking the Hamilton-Cayley theorem that a matrix satisfies its own
characteristic equation.  For example, let us consider the elementary case:
\be
({\cal M} - m \oneone)^2 = (u\Gamma_{012})^2 = u^2\quad
\Rightarrow \quad \mu = (m\pm u)\ ,\label{illustr}
\ee
where $\mu$ denotes the eigenvalues of ${\cal M}$.  Since $\Delta=4$ in
$D=11$, it is easy to see that $m=u$, and so the $32\times32$ matrix $\cal
M$ has 16 zero eigenvalues, for both the elementary membrane and solitonic
5-branes solutions.  We shall write the eigenvalues as
$2m(0_{16}, 1_{16})$, where the subscripts denote the
degeneracies.    Thus both the elementary and solitonic solutions break half
the supersymmetry and their mass/charge ratio is 1.  Although the
Hamilton-Cayley theorem provides a simple way to calculate the eigenvalues,
it is not always easy to determine the degeneracies by this approach.
Knowing the degeneracies is of particular interest for the supersymmetric
$p$-brane solutions, since it determines the fraction of the $D=11$
supersymmetry that is preserved.  For the supersymmetric cases, we calculate
the full set of 32 eigenvalues, using an explicit representation of
$32\times 32$ gamma matrices in $D=11$.

     The above analysis of supersymmetry can easily be generalised to lower
dimensions.  In fact the Nester form for maximal supergravity in any
dimension is just the Kaluza-Klein dimensional reduction of the
11-dimensional expression (\ref{nestor}).  For example, it follows from eqns
(\ref{vbd+1}), (\ref{transgress}) and (\ref{nestor}) that the Nester form
for type IIA supergravity in $D=10$ is given by
\bea
N^{\sst{AB}} &=& \bar \epsilon_1\Gamma^{\sst{ABC}} D_{\sst{C}} \epsilon_2 +
e^{-\ft34\phi}\, \bar \epsilon_1\Gamma_{10} \Big (\ft14 {\cal F}^{\sst{AB}} +
\ft18 \Gamma^{\sst{ABCD}} {\cal F}_{\sst{CD}} \Big)\epsilon_2\nonumber\\
&& + e^{\ft12\phi}\,\bar \epsilon_1\Gamma_{10}\Big( -\ft1{4}
\Gamma^{\sst{C}} F^{\sst{AB}}
{}_{\sst C} -\ft1{24} \Gamma^{\sst{ABCDE}}
F_{\sst{CDE}}\Big)\epsilon_2\label{bog10all}\\
&& + e^{-\ft14 \phi}\, \bar \epsilon_1
\Big( \ft18 \Gamma^{\sst{CD}} F^{\sst{AB}}{}_{
\sst{CD}} + \ft1{96} \Gamma^{\sst{ABCDEF}} F_{\sst{CDEF}}\Big)
\epsilon_2\ .\nonumber
\eea
The Nester form becomes increasingly complicated as we descend through the
dimensions, since more and more antisymmetric tensors are generated.  However,
for the purpose of studying the supersymmetries of $p$-brane solutions, some
simplifications can be made.   First, note that the dilaton factor for each
field strength is precisely the square root of the dilaton factor for the
kinetic term of the same field strength that appears in the Lagrangian.  In
fact it follows from (\ref{gensol}) that all these dilaton factors can be
set to unity since the Bogomol'nyi matrix we are considering is defined at
$r=\infty$.  As we showed above, in order to obtain the eigenvalues of a
Bogomol'nyi matrix, we do not need to decompose the $\Gamma$ matrices into
world-volume and transverse space factors.  Furthermore, we do not need to
decompose the 11-dimensional $\Gamma$ matrices into the product of
$D$-dimensional spacetime and compactified $(11-D)$-dimensional factors.
This greatly simplifies the discussion for lower dimensions.

     In order to present the general Bogomol'nyi matrix for arbitrary forms
and arbitrary dimensions, we first establish a notation for the Page charges
of the various field strengths:
\bea
{}&&F_4\qquad\quad F_3^i\quad\qquad F_2^{ij}\qquad\quad F^{ijk}_1
\quad\qquad {\cal F}_2^i\qquad\quad {\cal F}_1^{ij}\nonumber\\
{\rm elementary:}&& u\quad\qquad\,\,\, u_i\quad\qquad\,\, u_{ij}\quad\qquad
\phantom{v_{ijk}}\quad\qquad\,\, p_i\quad\qquad
\phantom{q_{ij}}\label{pagecharge}\\
{\rm solitonic:}&& v\qquad\quad \,\,\, v_i\quad\qquad \,\, v_{ij}\quad\qquad
\,\, v_{ijk}\quad\qquad \, q_i\quad\qquad\,\,\,\, q_{ij}\ ,\nonumber
\eea
where the elementary Page charges are given by $\ft{1}{\omega_{{\sst D}-n}}
\int_{S^{({\sst D}-n)}} *F_n$ and the solitonic Page charges are given by
$\ft{1}{4\omega_n} \int_{S^n} F_n$.  Note that there are no elementary
$p$-brane solutions for the 1-form field strengths.  We find that the
general Bogomol'nyi matrix in $D$ dimensions is given by
\bea
{\cal M} &=& m\oneone + u\, \Gamma_{012} + u_i\, \Gamma_{01i} +
\ft12 u_{ij}\, \Gamma_{0ij} +p_i \Gamma_{0i}\nonumber\\
&&+ v\,\Gamma_{\hat1\hat2\hat3\hat4\hat5} + v_i \,
\Gamma_{\hat1\hat2\hat3\hat4i}+\ft12 v_{ij}\, \Gamma_{\hat1\hat2\hat3ij}
+ \ft16 v_{ijk}\, \Gamma_{\hat1\hat2ijk} + q_i\, \Gamma_{\hat1\hat2\hat3 i}
+\ft12 q_{ij}\, \Gamma_{\hat1\hat2ij}\ ,\label{genbog}
\eea
where the first line contains the contributions for elementary solutions,
and the second line contains the contributions for solitonic solutions.  For
a given degree $n$ of antisymmetric tensor field strength, only the terms
with the corresponding Page charges, as indicated in (\ref{pagecharge}),
will occur.  As always, the indices $0, 1,\ldots$ run over the dimension of
the $p$-brane worldvolume, $\hat1,\hat2,\ldots$ run over the transverse
space of the $y^m$ coordinates, and $i,j,\ldots$ run over the dimensions
that were compactified in the Kaluza-Klein reduction from 11 to $D$
dimensions.  The mass per unit $p$-volume $m$ in (\ref{genbog}) arises from
the connection term in the covariant derivative in the Nester form, and it
is given by $m =\ft12 {\rm lim}_{r \rightarrow \infty}\, (B'-A')e^{-B}
r^{\tilde d + 1}$.  Thus the masses and Page charges of the solutions of
(\ref{genlag}) that we discussed in section 1 are given for all $D$ and $d$
by
\bigskip\bigskip

\centerline{
\begin{tabular}{|c|c|c|c|c|}\hline
  &\phantom{for}Elementary \phantom{for} &\phantom{for} Solitonic\phantom{for}
& Dyonic, $a^2=n-1$& Dyonic, $a^2=0$\\ \hline\hline
Mass $m$  & ${1\over 2 \sqrt\Delta}\lambda$ & ${1\over 2 \sqrt\Delta}\lambda$
& ${1\over 2\sqrt\Delta}\,(\lambda_1 +\lambda_2)$ &${1\over 2\sqrt\Delta}\,
\sqrt{\lambda_1^2+\lambda_2^2}$\\  \hline
Charge $P_{\rm ele}$ & $\ft14\lambda$  & $0$ & $\ft14 \lambda_1$  &$
\ft14\lambda_1 $ \\ \hline
Charge $P_{\rm sol}$ & $0$  & $\ft14\lambda $ & $\ft14\lambda_2$ &
$\ft14\lambda_2$ \\ \hline
\end{tabular}}
\bigskip

\centerline{Table 1: Mass and Page charges for $p$-brane solutions}

\bigskip

      As one can see from (\ref{illustr}), the existence of zero eigenvalues
of a Bogomol'nyi matrix depends on two factors: the $\Gamma$ matrix
structure and the $\Delta$ value of the solution.  Both of these two factors
are governed by the field strengths that participate in the solution.  In
the next three sections, we shall exhaustively search for $p$-brane solutions
using $n$-forms with $n=4,3,2,1$ in all dimensions $11\ge D \ge 4$.  In the
case of 4-forms and 3-forms we descend only to $D=8$ and $D=6$
respectively, since below these dimensions, the forms will be dualised to
forms of lower degree.   It will be understood in everything that
follows that we are always discussing $p$-brane solutions in the maximal
supergravity that is obtained from $D=11$ by dimensional reduction.

\section{$p$-branes for 4-form and 3-form field strengths}

\subsection{4-form field strengths}

    We first discuss $p$-brane solutions for 4-form field strengths in
$D\ge 8$ dimensions.  There is only one 4-form in each dimension $D$.
Therefore, in this case, there is no possibility of truncation of the dilatonic
scalar fields, and so there is a unique solution, with $\Delta = 4$.   When
it is elementary, it is a membrane; when it is solitonic, it is a
$(D-6)$-brane.  The Bogomol'nyi matrices for all these dimensions take the
same form (\ref{bog11}) as in 11 dimensions.   Thus the eigenvalues are
given by $m\pm u$ for the elementary cases and $m\pm v$ for the solitonic
cases.  It follows from eqn.\ (\ref{masscharge}) that all these solutions
preserve half of the 11-dimensional supersymmetry. Note that in $D=8$, one
could consider dyonic membrane solutions given by (\ref{dyonicsol1}).
However, as we discussed in section 2, such a solution is ruled out by the
constraint given in (\ref{ffaconstr}).  (It is worth remarking that one can
nevertheless construct a more general kind of dyonic solution where a 0-form
potential is in addition involved \cite{ilpt}.)

      In fact the above discussion applies also to the $p$-brane solutions
for $n$-forms of any other degree, if only one of the original field strengths
participates in the solution.   In other words, any $p$-brane solution with
$\Delta =4$ preserves half of the 11-dimensional supersymmetry, regardless
of the dimension $D$ and the degree $n$ of the form.   Thus in the following
discussion, we shall consider only the more complicated cases where more
than one field strength is involved, and hence $\Delta <4$.

\subsection{3-form field strengths}

    There are $(11-D)$ 3-forms in $D\ge 6$ dimensions, except in $D=7$ where
there is an extra 3-form coming from the dualisation of the 4-form.  The
3-forms are obtained from dimensional reduction of the 4-form in $D=11$.
Leaving aside the extra 3-form in seven dimensions, they can be labelled by
$F^{\sst{ABC}i}$, where $i$ runs over the $(11-D)$ compactified dimensions.
In the case of $D=7$, it turns out that the extra 3-form can be appended to
these by extending the range of the index $i$.   The dilaton vectors $\vec
a_i$ for the corresponding 3-form field strengths $F^{\sst{ABC}i}$, which
are given by (\ref{dilatonvec}), turn out to have the following dot products
\be
\vec a_i\cdot \vec a_j = 2 \delta_{ij} -\fft{2(D-6)}{D-2}
\ .\label{dot3form}
\ee

     Let us now suppose that there are $N$ 3-forms participating in a
$p$-brane solution, where $N$ is less than or equal to the number of
dilatonic scalar fields, {\it i.e.}\ $N\le (11-D)$.  It follows from the
first equation of (\ref{avaluesol}) that all these 3-forms are equal, {\it
i.e.}\ $F_\alpha^2=F^2/N$ for all $\alpha$, where $F$ is the canonically
normalised field strength that appears in the Lagrangian (\ref{genlag})
after setting the non-participating fields to zero.  Thus it follows from
eqn.\ (\ref{fequation1}) that
\be
a^2 =\Big (\fft1{N} \sum_{\alpha} \vec a_{\alpha}\Big)^2
=\fft{2}{N} - \fft{2(D-6)}{D-2}\ ,\label{avalue3f}
\ee
which implies that
\be
\Delta = 2 + \fft2{N}=4, \, 3, \, \fft83,\, \fft52,\,
\fft{12}5\ .\label{delta3f}
\ee
This implies that there are $(11-D)$ inequivalent $p$-brane solutions using
3-form field strengths in $D$ dimensions.  The elementary solutions are
strings, while the solitonic solutions are $(D-5)$-branes.

     In order to study the supersymmetry of these $p$-brane solutions we
consider their Bogomol'nyi matrices, given in (\ref{genbog}).  In $D=10,9,8$,
we therefore have
\bea
{\rm elementary:} && {\cal M} = m\oneone + u_i \Gamma_{01i}\ ,\quad
{\rm eigenvalues:}\quad  \mu = m \pm \sqrt{u_i u_i}\ ,\nonumber\\
{\rm solitonic:} && {\cal M} = m\oneone + v_i \Gamma_{\hat1\hat2\hat3\hat4 i}
\ ,\quad
{\rm eigenvalues:}\quad  \mu = m \pm \sqrt{v_i v_i}\ ,\label{bog3f1}
\eea
where we denote the elementary and solitonic Page charges by $u_i$ and
$v_i$ respectively, and the index $i$ runs over the compactified dimensions.
 As we discussed above, for a $p$-brane solution involving $N\le (11-D)$
field strengths, we have $F^2_\alpha = F^2/N$ for all $\alpha$, and hence
$u_\alpha = u/\sqrt{N}$ or $v_\alpha = v/\sqrt{N}$ where $u$ and $v$ are the
Page charges for the canonically normalised field strength $F$, {\it i.e.}\
$u = \fft14 \lambda = v$.  Thus the eigenvalues for both the elementary and
solitonic solutions are given by $\mu = \fft14 \lambda (\fft2{\sqrt{\Delta}}
\pm 1)$, and can be zero only when $\Delta = 4$, in which case half of the
supersymmetry is preserved.

     In $D=7$, there is an extra 3-form coming from the dualisation of the
4-form.  The Bogomol'nyi matrices are given by
\bea
{\rm elementary:} && {\cal M} = m\oneone + u_i \Gamma_{01i} + u \Gamma_{
\hat1\hat2\hat3\hat4\hat5}\ ,\quad
{\rm eigenvalues:}\quad  \mu = m \pm \sqrt{u_i u_i+ u^2}\ ,\nonumber\\
{\rm solitonic:} && {\cal M} = m\oneone + v_i \Gamma_{\hat1\hat2\hat3\hat4 i}
+ v \Gamma_{012}\ ,\quad
{\rm eigenvalues:}\quad  \mu = m \pm \sqrt{v_i v_i + v^2}\ ,\label{bog3f2}
\eea
where $u$ and $v$ denote the elementary and solitonic Page charges of the
dual of the 4-form field strength.   Note that, as we discussed for the
solutions, $u$ and $v$ play an equivalent r\^ole to $u_i$ and $v_i$
respectively in the Bogomol'nyi matrices, and therefore we can extend the
range of the index $i$ by 1.  Thus as in the cases of $D\ge 8$,  the
eigenvalues are given by $\mu = \fft14 \lambda (\fft2{\sqrt{\Delta}}
\pm 1)$.  Again the supersymmetry is completely broken unless $N=1$ and
hence $\Delta = 4$, in which case half of the supersymmetry is preserved.

     In $D=6$,  the Bogomol'nyi matrices are given by
\bea
{\cal M} &=& m\oneone + u_i \Gamma_{01i} + v_i \Gamma_{\hat1\hat2\hat3
\hat4 i}\ ,\nonumber\\
{\rm eigenvalues:}&& \mu = m \pm \sqrt{(u_i \pm v_i)^2}\ .
\label{bog3f3}
\eea
(Note that owing to the more complex $\Gamma$-matrix structure, the
characteristic equation for ${\cal M}$ is quartic here rather than
quadratic, giving rise to the two independent $\pm$ signs.)  The elementary
and solitonic solutions correspond to $v_i = 0$ and $u_i = 0$ respectively.
For these types of solutions, the supersymmetry properties are identical to
the case when $D\ge 7$, namely all solutions break all the supersymmetry
except when $\Delta = 4$, in which case they preserve half the
supersymmetry.

    In $D=6$, since the dimension is twice the degree of the 3-form field
strengths, we can also construct dyonic string solitons, where the field
strengths involve both elementary and solitonic contributions.  There could
in principle be two different types of dyonic solution. In the first type,
some field strengths would be purely elementary (with electric charge) while
others would be purely solitonic (with magnetic charge).  However, such
configurations do not satisfy the constraint given in (\ref{ffaconstr}).
In the second type of solution, the canonical field strength has
both electric and magnetic charges, and hence all the participating field
strengths have the same ratio of the two charges.   This configuration does
satisfy the constraint in (\ref{ffaconstr}).  Thus we can have such dyonic
solutions with $N=1,2,3,4$ and 5 field strengths, and the correponding
values of $\Delta$ are given by (\ref{delta3f}).  In general, the equations
of motion reduce to the two second-order differential equations
(\ref{phiadiff}). When $N=1$, and hence $\Delta=4$, we have a simple
solution given by (\ref{dyonicsol1}).  The mass per unit length of this
dyonic string is $m=\ft14(\lambda_1 +\lambda_2)$, and the Page charges are
$u=\ft14\lambda_1$ and $v=\ft14 \lambda_2$.  The eigenvalues of the
Bogomol'nyi matrix are given by
\be
\mu = m\pm u \pm v = \{0_8, (\ft12 \lambda_1)_8,
(\ft12 \lambda_2)_8, (\ft12 (\lambda_1 + \lambda_2))_8 \}\ ,\label{extra1}
\ee
where, as usual, the subscripts on the eigenvalues indicate their
degeneracies.  Thus the solution preserves $\ft14$ of the supersymmetry
\cite{dfkr}.  When either $\lambda_1=0$ or $\lambda_2=0$, the solution
reduces precisely to the previously-discussed purely solitonic and purely
elementary solutions, which preserve $\ft12$ of the supersymmetry.  When
$\lambda_1=\lambda_2$, in which case the field strength becomes self-dual
and the dilaton vanishes, the solution is in fact precisely equivalent to
the self-dual string in $D=6$ self-dual supergravity, which we shall discuss
below.  When $\lambda_1 =-\lambda_2$, the field strength is anti-self-dual,
and we have a massless string which preserves $\ft12$ of the supersymmetry;
however, the eigenvalues, given by (\ref{extra1}), for such a solution are
not positive semi-definite.   In this case, the dilaton field does not
vanish, and hence the solution is distinct from the anti-self-dual string in
$D=6$ anti-self-dual gravity.  It is worth remarking that the eigenvalues
(\ref{extra1}) for these dyonic solutions of the second type are quite
different from those for all the solutions we have discussed previously.
In those cases, the eigenvalues are non-negative as long as the mass per
unit $p$-volume is positive.  However, for the dyonic solutions of the
second type, one can easily see that the eigenvalues (\ref{extra1}) of the
Bogomol'nyi matrices can take both signs, even when the mass is positive.

     In the above discussion, we saw that the field strength of the solution
could be chosen to be either self-dual or anti-self-dual.  In fact, one can
alternatively impose a self-dual (or anti-self-dual) condition on a 3-form
field strength so as to truncate the supergravity theory itself to an
(anti)-self-dual theory \cite{romans}. In this case, the dilatonic fields
are all consistently truncated from the theory. This gives rise to a
self-dual (or anti-self-dual) string soliton with $\Delta =2$.  The metric
is given by (\ref{dyonicsol2}) with $\lambda_1=\lambda_2=\lambda$. The mass
per unit length is given by $m=\ft12 \sqrt{(\lambda_1^2 +
\lambda_2^2)/\Delta} = \ft12 \lambda$; the Page charges of the solution
comprise an electric charge $u$ and a magnetic charge $v$, with $u=v=\ft14
\lambda$ ($v=-\ft14 \lambda$ for the anti-self-dual case). The
eigenvalues of the Bogomol'nyi matrix are given by $\mu = \fft{\lambda}{4}
(2\pm 1\pm1)= m \{0_8,1_{16}, 2_8\}$, and so a quarter of the $D=11$
supersymmetry is preserved in this (anti)-self-dual case.  Note that the
mass per unit $p$-volume of the self-dual (anti-self-dual) solution in the
previous paragraph is given by $m=u + v$, whilst the mass of these
solution in $D=6$ self-dual (anti-self-dual) supergravity is given by
$m=\sqrt{u^2 + v^2}$.

\section{$p$-branes for 2-form field strengths}

      The structure of the 2-form field strengths is much complicated than
that for 4-forms and 3-forms.  There are two sources of 2-form fields
strengths, namely the vielbein and the dimensional reduction of the 4-form
in $D=11$.  Furthermore, in $D=6$, there is an additional 2-form coming from
the dualisation of the 4-form, and in $D=5$, there are six additional
2-forms coming from the dualisation of the 3-forms.  Thus we shall discuss
$D\ge 7$, $D=6$, $D=5$ and $D=4$ separately.  The elementary solutions are
particles, while the solitonic solutions are $(D-4)$-branes.

\subsection{$D\ge 7$}

     When $D\ge 7$, there are $(11-D)$ 2-forms ${\cal F}_{\sst{MN}}^{(i)}$
coming from the vielbein, and $\ft12 (11-D)(10-D)$ 2-forms $F_{\sst{MN}ij}$
coming from the 4-form field strength, where $i$ runs over the compactified
dimensions. We are denoting the dilaton vectors by $\vec b_i$ and $\vec
a_{ij}$ respectively; they are given in eqn.\ (\ref{dilatonvec}).  It
follows from eqn.\ (\ref{gfdot}) that they satisfy
\bea
\vec b_i\cdot \vec b_j &=& 2\delta_{ij} + \fft{2}{D-2}\ ,\nonumber\\
\vec b_i \cdot \vec a_{jk} &=& -2 \delta_{ij} - 2\delta_{ik} +\fft2{D-2}
\ ,\qquad j < k\ , \label{dot2form1}\\
\vec a_{ij}\cdot \vec a_{kl} &=& 2 \delta_{ik} + 2 \delta_{jl} +
2 \delta_{il} + 2 \delta_{jk} - \fft{2(D-3)}{D-2}\ ,
\qquad i < j, \, k < l.\nonumber
\eea

       The Bogomol'nyi matrix for the elementary case is
${\cal M} = m\oneone + p_i \Gamma_{0i} + \ft12 u_{ij} \Gamma_{0ij}$.  Its
characteristic equation is quartic, and the eigenvalues are
\bea
\mu &=& m \pm \sqrt{p_i p_i + \ft12 u_{ij}u_{ij} \pm
2\sqrt{u_{ij} p_i u_{kj} p_k}}\ ,\qquad\qquad\qquad\qquad
D=10,9,8\nonumber\\
\mu &=& m \pm \sqrt{p_i p_i + \ft12 u_{ij}u_{ij} \pm
\sqrt{\ft1{16} (\epsilon_{ijkl} u_{ij} u_{kl} )^2
+ 4 u_{ij} p_i u_{kj} p_k }}\ ,\qquad D=7\ .\label{bog2f1}
\eea
For the solitonic case, the analysis is analogous, and the
eigenvalues are given by (\ref{bog2f1}) with the elementary Page
charges $p_i$ and $u_{ij}$ replaced by the solitonic Page charges $q_i$ and
$v_{ij}$.

     Using eqns (\ref{dot2form1}) and (\ref{bog2f1}), we can obtain all the
$p$-brane solutions for 2-form field strengths in $D\ge 7$, and study their
supersymmetry properties.  As we discussed previously, the $p$-brane
solutions involving only one original 2-form have $\Delta =4$, and preserve
half of the supersymmetry.  For $N=2$ participating 2-forms, which can
happen in $D \le 9$, there are two inequivalent solutions for each $D$, with
$\Delta =3$ and $\Delta =2$.  For each value of $\Delta$, there may be more
than one choice of field strengths to achieve it; however, as long as the
ratio of field strengths is the same for the different choices, we expect
that they give rise to identical eigenvalues in the Bogomol'nyi matrix, and
hence the same supersymmetry property.  Thus we shall present only one
example for each inequivalent solution:
\bea
\Delta = 3\ , && p_1 = p_2 = \ft{\lambda}{4\sqrt2}\ ,\qquad
\mu = \ft{\lambda}{4} (\fft2{\sqrt3} \pm 1)\ ,\nonumber\\
\Delta = 2\ , && p_1 = u_{12} = \ft{\lambda}{4\sqrt2}\ ,\qquad
\mu = \ft{\lambda}{4\sqrt2} (2\pm 1\pm1)\ . \label{deltad71}
\eea
We see that the $p$-brane solutions with $\Delta =3$ break all the
supersymmetry, and the $p$-brane solutions with $\Delta = 2$ preserve a
quarter of the supersymmetry, with the eigenvalues given by
$m\{0_{8}, 1_{16}, 2_{8}\}$.

     For $N=3$ field strengths, which can happen in $D\le 8$, there are also
two inequivalent solutions for each $D$, namely
\bea
\Delta = \ft83\ ,&& \{p_1,p_2,p_3\} = \ft{\lambda}{4\sqrt3}\{1,1,1\}\ ,
\qquad \mu = \ft{\lambda}{4} ( \sqrt{\ft32} \pm 1)\ ,\nonumber\\
\Delta = \ft{12}{7}\ ,&& \{p_1, p_2, u_{12}\} =\ft{\lambda}{4\sqrt7}
\{\sqrt2,\sqrt2,\sqrt3\}\ ,\qquad
\mu = \ft{\lambda}{4} (\sqrt{\ft73} \pm \sqrt{ 1 \pm \ft{4\sqrt3}7})\ .
\label{deltad72}
\eea
These solutions break all the supersymmetry.

      For $N=4$ field strengths, which can happen in $D\le 7$,  there are 3
inequivalent solutions, given by
\bea
\Delta&=& \ft52\ ,\quad \{p_1,p_2,p_3,p_4\} = \ft{\lambda}{8}\{1,1,1,1\}\ ,
\quad \mu= \ft{\lambda}{4} (\sqrt{\ft85}\pm 1)\ ,\nonumber\\
\Delta &=& \ft53\ ,\qquad \{p_1,p_2,u_{12},u_{13}\} =\ft{\lambda}{4\sqrt6}
\{\sqrt2,1,\sqrt2,1\}\ ,\quad \mu = \ft{\lambda}{4} (\sqrt{\ft{12}{5}}
\pm \sqrt{1\pm \ft{\sqrt8}{3}} )\ ,
\label{deltad73}\\
\Delta &=& \ft85\ ,\quad \{p_1, u_{12}, u_{13}, u_{14}\} =
\ft{\lambda}{4\sqrt5}\{ \sqrt2,1,1,1\}\ ,\quad
\mu = \ft{\lambda}{4}(\sqrt{\ft52} \pm \sqrt{1\pm \ft{2\sqrt6}{5}})\ .
\nonumber
\eea
All these solutions, which break all the supersymmetry, satisfy the
constraints implied by both ${\cal L}_{\sst{FFA}}$ and the Chern-Simons
modifications that we discussed in section 2. Note that in order to get
$\Delta = \ft53$, we can also have $\{ p_1, p_2, p_3, u_{12}\} =
\ft{\lambda}{4\sqrt6} \{\sqrt2, \sqrt2, \sqrt{-1}, \sqrt3\}$. The
eigenvalues are given by $\mu=\ft{\lambda}{4} (\sqrt{\ft{12}5} \pm \sqrt{ 1
\pm \ft2{\sqrt3}})$.  In this case, one of the field strengths is imaginary
and $\mu$ includes complex values.  The solution is unphysical. From now on,
we shall discard these types of solutions.

\subsection{$D=6$}

      In $D=6$, there are five 2-forms ${\cal F}_{\sst{MN}}^{(i)}$ coming
from the vielbein,  ten $F_{\sst{MN}ij}$ coming from dimensional reduction
of the 4-form and an extra 2-form coming from the dualisation of the 4-form.
 We denote the corresponding dilaton vectors by $\vec b_i$, $\vec a_{ij}$
and $\vec a$.  Note that $\vec a$ is given by eqn.\ (\ref{dilatonvec}) but
with its sign reversed because of the dualisation. The dot products of these
vectors are given by eqn.\ (\ref{dot2form1}), together with
\be
\vec a \cdot \vec a = \ft52\ ,\qquad\vec a \cdot \vec b_i = -\ft32\ ,
\qquad \vec a \cdot \vec a_{ij} = \ft12\ ,\qquad i< j \ ,\label{dot2form2}
\ee
The Bogomol'nyi matrix for the elementary case is given by ${\cal M} =
m\oneone + p_i \Gamma_{0i} + \ft12 u_{ij} \Gamma_{0ij} + u^*\,
\Gamma_{\hat1\hat2\hat3\hat4\hat5}$, where $u^*$ is the Page charge for the
extra 2-form coming from the dualisation of the 4-form.  The eigenvalues are
given by
\bea
\mu &=& m \pm \sqrt{p_i p_i + u^{*2} + \ft12 u_{ij}u_{ij} \pm \sqrt{X}}\ ,
\nonumber\\
X &=& 4 u_{ij} p_j u_{ik} p_k + 4 p_i p_i u^{*2} +
\ft12 (u_{ij}u_{ij} )^2
-u_{ij} u_{jk} u_{kl} u_{li}  + \epsilon_{ijklm} u_{ij} u_{kl} p_m u^*
\ .\label{bog2f2}
\eea
The eigenvalues for the solitonic case take the same form, with the
elementary Page charges replaced by the corresponding solitonic Page
charges.

     Now we shall consider $p$-brane solutions with $N\le 5$ field
strengths.   When $N=1,2,3$, the solutions are identical to the cases when
$D=7$.  When $N=4$, in addition to the 3 inequivalent solutions that we
discussed in $D=7$, there is an extra inequivalent solution, given by
\be
\Delta = \ft32\ ,\quad \{p_1, u_{12}, u_{13}, u_{45}\} =
\ft{\lambda}{8}\{1,1,1,1\}
\ ,\quad \mu = \ft{\lambda}{4} (\sqrt{\ft83} \pm \sqrt{1\pm 1})\
.\label{deltad61}
\ee
The solution is not supersymmetric.

      When $N=5$, there are three inequivalent solutions, given by
\bea
\Delta &=& \ft{12}5 \ ,\qquad \{p_1,p_2,p_3,p_4,p_5\}
=\ft{\lambda}{4\sqrt5} \{1,1,1,1,1\}\ ,\nonumber\\
\Delta &=& \ft85 \ ,\qquad \{p_1, u_{12},u_{14},u_{23},u_{45}\} =
\ft{\lambda}{4\sqrt5} \{1,1,1,1,1\}\ ,\label{deltad62}\\
\Delta &=& \ft{20}{13}\ ,\qquad \{u_{12}, u_{13}, u_{14}, u_{15}, p_1\}
=\ft{\lambda}{4\sqrt{13}}\{\sqrt2,\sqrt2,\sqrt2,\sqrt2,\sqrt5\}\ ,\nonumber
\ .
\eea
These have eigenvalues $\mu=\ft{\lambda}{4}(\sqrt{\ft53} \pm 1)$,
$\mu = \ft{\lambda}{4} (\sqrt{\ft52} \pm \sqrt{1 \pm \ft{2}{\sqrt5}})$ and
$\mu = \ft{\lambda}{4}(\sqrt{\ft{13}5} \pm
\sqrt{1 \pm \ft{4\sqrt{10}}{13}})$ respectively, and so none of these
solutions is supersymmetric.

\subsection{$D=5$}

      In $D=5$, there are six 2-forms ${\cal F}_{\sst{MN}}^{(i)}$ from the
vielbein,  fifteen $F_{\sst{MN}ij}$ coming from dimensional reduction of the
4-form and six extra 2-forms coming from the dualisation of the 3-forms
$F_{\sst{MNP}i}$.  We denote the corresponding dilaton vectors by $\vec
b_i$, $\vec a_{ij}$ and $\vec a_i$.  Note that the vectors $\vec a_i$ are
given by eqn.\ (\ref{dilatonvec}) but with their signs reversed owing to the
dualisation. The dot products of these vectors are given by eqn.\
(\ref{dot2form1}), together with
\bea
\vec a_i \cdot \vec a_j &=& = 2 \delta_{ij} +\ft23\ ,\qquad
\vec a_i\cdot \vec b_j = 2 \delta_{ij} -\ft43\ ,\nonumber\\
\vec a_i \cdot \vec a_{jk} &=&  -2 \delta_{ij} - 2 \delta_{ik} +\ft23
 \ ,\qquad j < k\ ,\label{dot2form3}
\eea
The Bogomol'nyi matrix for the elementary case is given by
\be
{\cal M} = m\oneone + u_i \Gamma_{0i} + \ft12 u_{ij} \Gamma_{0ij} + u^*_i
\, \Gamma_{\hat1\hat2\hat3\hat4 i}\ , \label{bog2f3}
\ee
where $u^*_i$ are the Page charges for the extra six 2-forms coming from the
dualisation of the 3-forms.  The general formula for the eigenvalues for
this matrix is too complicated to present. We shall instead just give the
eigenvalues for the explicit inequivalent solutions that we find.  Since the
eigenvalues of the Bogomol'nyi matrix for the solitonic case take the
identical form, we shall not discuss them further.

     In $D=5$, there can be $p$-brane solutions with $N\le 6$ participating
field strengths.  The cases with $N=1,2,4$ are precisely the same as in $D=6$,
but for $N=3,5$ there is an additional solution in each case.  Together
with five solutions for $N=6$, the full set of new solutions in $D=5$ is
\bea
\Delta &=& \ft43\ ,\qquad \{ u_{12},u_{34},u_{56}\} =
\ft{\lambda}{4\sqrt3}\{1,1,1\}\ ,\nonumber\\
\Delta &=& \ft{24}{17}\ ,\qquad \{u_{12},u_{13}, u_{45}, u_{56}, u_{46}
\} =\ft{\lambda}{8\sqrt5}\{2,2,\sqrt3,\sqrt3,\sqrt3\}\ ,\nonumber\\
\Delta &=& \ft73\ ,\qquad \{p_1,p_2,p_3,p_4,p_5,p_6\}
=\ft{\lambda}{4\sqrt6} \{1,1,1,1,1,1\}\ ,\nonumber\\
\Delta &=& \ft32\ ,\qquad \{p_1, u_{12},u_{13},u_{14},u_{15},u_{16}\} =
\ft{\lambda}{8\sqrt2} \{ \sqrt3,1,1,1,1,1\}\ ,\nonumber\\
\Delta &=& \ft{15}{11}\ ,\qquad \{ p_1, p_{6}, u^*_{2}, u^*_{3},
u^*_{4}, u^*_{5} \} =
\ft{\lambda}{4\sqrt22} \{\sqrt5,\sqrt5,\sqrt3,\sqrt3,\sqrt3,\sqrt3\}
 \ ,\label{deltad5}\\
\Delta &=& \ft75\ ,\qquad \{u_{26},u_{25}, p_1,p_2, u_{13},u_{14}\} =
\ft{\lambda}{4\sqrt10} \{ \sqrt2, \sqrt2, 1,1, \sqrt2, \sqrt2\}
\ ,\nonumber\\
\Delta &=& \ft43\ ,\qquad \{u_{12}, u_{13}, u_{23}, u_{45}, u_{46},
u_{56}\} =\ft{\lambda}{4\sqrt6}\{1,1,1,1,1,1\}\ .\nonumber
\eea
All these solutions, except for the case of $\Delta =\ft43$ with three field
strengths, break all the supersymmetry.  From now on, we shall not present
the eigenvalues for the non-supersymmetric cases.   The eigenvalues for the
case of $\Delta=\ft43$ with three field strengths are $\ft23 m
\{0_4,1_{12},2_{12},3_{4}\}$.  Thus the $p$-brane solution with $\Delta =
\ft43$ preserves $\ft18$ of the supersymmetry.  Note that there is also a
case with $\Delta =\ft43$ for six field strengths, which does not give rise
to supersymmetric $p$-brane solutions.

\subsection{$D=4$}

      In $D=4$, there are a total of twenty-eight 2-form field strengths:
seven ${\cal F}_{\sst{MN}}^{(i)}$ from the vielbein,  and twenty-one
$F_{\sst{MN}ij}$ coming from dimensional reduction of the 4-form. The
corresponding dilaton vectors are denoted by $\vec b_i$ and $\vec a_{ij}$.
It turns out that the dot products of these vectors, given by
(\ref{dilatonvec}) and (\ref{gfdot}), are such that we can extend the $i$
index to include the value 8, and define $\vec a_{i 8} = -\vec b_i$.  Then
the dot products are given by the single formula
\be
\vec a_{ij} \cdot \vec a_{kl} = 2 \delta_{ik} + 2\delta_{jl}
+ 2 \delta_{il} + 2 \delta_{jk} - 1\ ,\qquad i<j\ , k<l\ ,
\label{dot2form4}
\ee
Although the dilaton vectors can be treated equally by this extension of the
index range, the corresponding field strengths do not play equivalent
r\^oles.

      In $D=4$,  since the dual of a 2-form is again a 2-form, in
addition to the previous linear combinations of original field strengths
we can also combine field strengths and the duals of other field strengths
to obtain a $p$-brane solution.   Such a phenomenon occurs also in $D=8$ for
the 4-form and $D=6$ for the 3-forms, as we discussed in the previous
section.  In terms of the original undualised fields, this corresponds to
making an elementary ansatz on some of the fields while making a solitonic
ansatz on the remainder.  In order to obtain such {\it dyonic} solutions,
we first dualise certain field strengths,  and then use the previously
discussed procedure to obtain the Lagrangian (\ref{genlag}) by taking a
linear combination of these together with other undualised field strengths.
We then make a purely elementary or purely solitonic ansatz on this
resulting field strength.  The corresponding Page charges of the dualised
field strengths are magnetic and electric respectively, precisely opposite
to those of the undualised field strengths.  As we discussed in section 4.2,
these are called dyonic solutions of the first type.  We shall discuss
dyonic solutions of the second type at the end of the section.

    The Bogomol'nyi matrix for all the above mentioned $p$-brane solutions
takes the form
\be
{\cal M} = m\oneone + p_i \Gamma_{0i} + \ft12 u_{ij} \Gamma_{0ij} +
q_i \Gamma_{\hat1\hat2\hat3i}
+ \ft12 v_{ij} \Gamma_{\hat1\hat2\hat3 ij}\ . \label{bog2f4}
\ee
An elementary ansatz for an undualised field strength contributes a $p_i$ or
$u_{ij}$ electric charge, whilst an elementary ansatz for a dualised field
strength contributes a $q_i$ or $v_{ij}$ magnetic charge.  When we present
the solutions later in the section, we shall denote such a $q_i$ or $v_{ij}$
charge by $p_i^*$ or $u_{ij}^*$. The situation for a solitonic ansatz is the
converse of this.

     Since there are a total of seven dilatonic scalars in $D=4$, there can
be $N\le 7$ participating field strengths.   In general, there can be either
purely electric (or purely magnetic), or else mixed dyonic solutions. For
$N=1$, the solutions are necessarily either electric or magnetic, with
$\Delta = 4$, analogous to the cases in $D\ge 5$ that we have discussed
previously.  All these solutions preserve half the supersymmetry.  For
$N=2$,  there are two different values of $\Delta$ arising from the
solutions, namely $\Delta = 3$ and $\Delta = 2$.  Both of these values can
be achieved by purely electric (or purely magnetic) solutions.  They can
also be achieved by mixed dyonic solutions.   For higher numbers of field
strengths, starting at $N=4$, there are cases that can only be achieved by
mixed dyonic configurations; we shall refer to these as intrinsically dyonic
solutions.  For the purpose of presenting the results,  we shall continue
with the policy of not giving again solutions that can be obtained from a
higher dimension by dimensional reduction, {\it i.e.}\ solutions that have
the same $\Delta$ values and the same ratios of participating field
strengths.   For the new solutions that we do list, we shall favour the
purely electric (or purely magnetic) choices where possible, and present
dyonic solutions only when they are intrinsically dyonic.  We find that the
new solutions in $D=4$ are as follows
\bea
\Delta = 1\ ,&& \{ u_{57}, u_{46}, u_{23}, p_1^* \} =\ft{\lambda}{8}
\{1,1,1,1\}\ ,\nonumber\\
\Delta =\ft65\ ,&& \{u_{57}, u_{47}, u_{36}, u_{12} \} =
\ft{\lambda}{4\sqrt{10}} \{ \sqrt2,\sqrt2,\sqrt3,\sqrt3 \}\ ,\nonumber\\
\Delta = \ft{12}{11}\ ,&& \{ u_{57}^*, u_{47}^*, u_{37}, u_{27}, u_{45}\}
=\ft{\lambda}{4\sqrt{11}} \{ \sqrt2,\sqrt2,\sqrt2,\sqrt2,\sqrt3 \}\ ,
\nonumber\\
\Delta = \ft{8}{7}\ ,&& \{ u_{57}, u_{47}, u_{36}, u_{45}, u_{12} \}
=\ft{\lambda}{4\sqrt7}\{ 1,1,\sqrt2,1,\sqrt2\}\ ,\nonumber\\
\Delta = \ft{20}{17}\ ,&& \{u_{57}, u_{47}, u_{56}, u_{36}, u_{12} \}
=\ft{\lambda}{4\sqrt{17}}\{ \sqrt2,2,\sqrt2,2,\sqrt5\}\ ,\nonumber\\
\Delta = \ft{10}{9}\ ,&&\{u_{57}^*, u_{47}, u_{37}, u_{27}, u_{17}, u_{56}\}
=\ft{\lambda}{12\sqrt2}\{\sqrt5,\sqrt2,\sqrt2,\sqrt2,
\sqrt2,\sqrt5\}\ ,\nonumber\\
\Delta = \ft{24}{13}\, && \{ u^*_{57}, u^*_{47}, u_{37}, u_{27}, u_{17},
u_{45}\} = \ft{\lambda}{4\sqrt{23}}\{2,2,\sqrt3,\sqrt3,\sqrt3,\sqrt6\}\ ,
\nonumber\\
\Delta=\ft{8}{7}\ ,&&\{ u_{57}, u_{47}, u_{56}, u_{36}, u_{34}, u_{12}\}
=\ft{\lambda}{4\sqrt7}\{1,1,1,1,1,\sqrt2\}\ ,\nonumber\\
\Delta=\ft{7}{6}\ ,&&\{u_{57}, u_{47}, u_{37}, u_{56}, u_{24}, u_{13}\}
=\ft{\lambda}{8\sqrt3}\{1,1,1,\sqrt3,\sqrt3,\sqrt3\}\ ,\nonumber\\
\Delta=\ft{7}{6}\ ,&&\{u_{57}, u_{47}, u_{56}, u_{36}, u_{24}, u_{13}\}
=\ft{\lambda}{8\sqrt3}\{\sqrt2,1,\sqrt2,1,\sqrt3,\sqrt3\}\ ,\label{nsd4}\\
\Delta=\ft32\ ,&&\{ u_{57}, u_{47}, u_{37}, u_{27}, u_{17}, p_1\}
=\ft{\lambda}{8\sqrt2}\{1,1,1,1,1,\sqrt3\}\ ,\nonumber\\
\Delta = \ft73\ ,&&\{ u_{57}, u_{47}, u_{37}, u_{27}, u_{17}, u_{67}\}
=\ft{\lambda}{4\sqrt6}\{1,1,1,1,1,1\}\ ,\nonumber\\
\Delta = \ft{60}{59}\ ,&&\{ u_{57}^*, u_{47}, u_{37}, u_{27}, u_{17}, u_{56},
u_{67}\}=\ft{\lambda}{4\sqrt{59}}\{\sqrt{10},\sqrt6,\sqrt6,\sqrt6,\sqrt6,
\sqrt{15},\sqrt{10} \}\ ,\nonumber\\
\Delta = \ft{12}{11}\ ,&&\{ u_{57}^*, u_{47}, u_{37}, u_{27}, u_{17}, u_{56},
p_{6}\}=\ft{\lambda}{4\sqrt{11}}\{\sqrt{3},1,1,1,1,
\sqrt{3},1 \}\ ,\nonumber\\
\Delta = \ft{8}{7}\ ,&&\{ u_{57}, u_{47}, u_{56}, u_{36}, u_{24}, u_{13},
u_{12}\}=\ft{\lambda}{4\sqrt{7}}\{1,1,1,1,1,1,1 \}\ ,\nonumber\\
\Delta = \ft{40}{31}\ ,&&\{ u_{57}, u_{47}, u_{37}, u_{26}, u_{16}, u_{12},
p_{7}\}=\ft{\lambda}{4\sqrt{31}}\{\sqrt5,\sqrt5,\sqrt5,2,2,2,2 \}\ ,\nonumber\\
\Delta = \ft{28}{19}\ ,&&\{ u_{57}, u_{47}, u_{37}, u_{27}, u_{17}, u_{67},
p_{7}\}=\ft{\lambda}{4\sqrt{19}}\{\sqrt2,\sqrt2,\sqrt2,\sqrt2,\sqrt2,
\sqrt2,\sqrt7 \}\ ,\nonumber\\
\Delta = \ft{16}{7}\ ,&&\{ u_{57}, u_{47}, u_{45}, p_{1}, p_{2}, p_{3},
p_{6}\}=\ft{\lambda}{12}\{1,1,1,1,1,1,1 \}\ ,\nonumber
\eea
where a star on a Page charge indicates that the associated field strength
is dualised; these cases are the intrinsically dyonic solutions.  Note that
all of them satisfy the constraints implied by ${\cal L}_{\sst{FFA}}$ and
the Chern-Simons modifications that we discussed in section 2.  In fact,
this is the first time that we encounter cases where ostensible solutions
are actually ruled out by the constraints.  For example, the following
configurations
\bea
\Delta = \ft{15}{14}\ ,&&\{ p_1^*,p_2^*,p_3, u_{45}, u_{23}, u_{36}\}
=\ft{\lambda}{8\sqrt7}\{\sqrt6,\sqrt3,\sqrt5,\sqrt5,\sqrt3,\sqrt6\}
\ ,\nonumber\\
\Delta = \ft{40}{39}\ ,&&\{ p_1^*, p_2^*, p_3, u_{45}, u_{56}, u_{23},
u_{37}\}=\ft{\lambda}{4\sqrt{39}}\{\sqrt{8},\sqrt4,\sqrt5,\sqrt5,\sqrt5,
\sqrt{4},\sqrt{8} \}\ ,\nonumber\\
\Delta = \ft{28}{27}\ ,&&\{ p_{1}, p_{2}^*, p_{3}^*, u_{45}, u_{46}, u_{23},
u_{17}\}=\ft{\lambda}{12\sqrt{3}}\{\sqrt{2},2,2,2,2,
\sqrt{2},\sqrt{7} \}\ ,\nonumber\\
\Delta = \ft{24}{23}\ ,&&\{ p_{1}^*, p_{2}^*, u_{34}, u_{35}, u_{26}, u_{67},
u_{17}\}=\ft{\lambda}{4\sqrt{23}}\{\sqrt{3},\sqrt3,2,2,\sqrt3,
\sqrt{3},\sqrt{3} \}\ ,\\
\Delta = \ft{16}{15}\ ,&&\{ p_{1}^*, u_{23}, u_{24}, u_{25}, u_{67}, u_{36},
u_{14}\}=\ft{\lambda}{4\sqrt{15}}\{\sqrt{3},1,1,\sqrt3,\sqrt3,
\sqrt{2},\sqrt{2} \}\ ,\nonumber\\
\Delta = \ft{4}{3}\ ,&&\{ u_{57}, u_{47}, u_{37}, u_{27}, u_{17},
u_{67}^*,p_{7}\}=\ft{\lambda}{12}\{1,1,1,1,1,\sqrt{2},\sqrt{2} \}\ ,\nonumber
\eea
would be solutions were it not for the fact that they do not satisfy the
constraint (\ref{csconstr}) implied by the Chern-Simons modifications to the
field strengths.  This is because in each case, there is an index in common
between an undualised 2-form and a dualised 2-form, one coming from the
vielbein and the other coming from the field strength $F_4$ in $D=11$.

     Amongst the new solutions in (\ref{nsd4}), all but the first one, which
has $\Delta =1$, break all the supersymmetry.  We find that the eigenvalues
of the Bogomol'nyi matrix ${\cal M}$ for the first solution are given by
$m (0_{4},1_{24},2_{4})$, and thus it preserves $\ft18$ of the
supersymmetry.  Together with the solutions that can be obtained from higher
dimensions by dimensional reduction, there are a total of four inequivalent
supersymmetric particle solutions in $D=4$. They have $\Delta = \ft4{N}$ for
$N=1,2,3,4$, corresponding to $a=\sqrt3, 1, \ft1{\sqrt3}, 0$. (The analogous
solutions with the same values of $a$ were also obtained \cite{dj} in the
four-dimensional $N=4$ heterotic string.) For $N=1,2,3$, the solutions can
be purely electric or purely magnetic or mixed dyonic. They preserve
$2^{-N}$ of the original $D=11$ supersymmetry. For $N=4$, the solutions are
intrinsically dyonic and preserve $\ft18$ of the supersymmetry. Although the
solutions with $N=3$ and $N=4$ both preserve $\ft18$ of the supersymmetry,
the degeneracies of the non-zero eigenvalues of their Bogomol'nyi matrices
are different. The solutions for $N=1,2,3$ and $4$ first appear at $D=10, 9,
5$ and 4 dimensions respectively.  They correspond to what are called
stainless super $p$-branes in \cite{lpss}, in the sense that they cannot
``oxidise'' to isotropic $p$-brane solutions in higher dimensions by the
inverse of the dimensional reduction procedure.

     So far we have considered either pure elementary or pure solitonic
solutions, or else mixed dyonic $p$-brane solutions of the first type, where
some of the field strengths are purely elementary and the others are purely
solitonic.  We now turn to dyonic solutions of the second type, where
the canonical field strength has both elementary and solitonic
contributions, and hence each participating field strength has the same
ratio of electric and magnetic charges.  As in the first type of dyonic
solution,
here too the constraint given in (\ref{ffaconstr}) must also be satisfied.
This is automatic for 2-forms coming from the vielbein; however, if two or
more of the
2-forms $F_2^{ij}$ coming from $\hat F_4$ in $D=11$ are involved, then these
must each have an index in common with all the others in order to satisfy
the constraint.  In fact all the
previously-listed examples of 2-form field
strength configurations can also be used to contruct these
dyonic solutions of the second type.  However, as we discussed in section 1,
the equations of motion only admit simple solutions when $a=0$ or 1,
corresponding to $\Delta=1$ or 2.  The metrics of these two dyonic particles
are given by
(\ref{dyonicsol1}) and (\ref{dyonicsol2}) respectively.  The corresponding
masses, Page charges and the eigenvalues of Bogomol'nyi matrices
are given by
\bea
\Delta=1: && m=\ft12   \lambda_{12}\ ,\qquad
\{u_{57}, u_{46}, u_{23}, p_1^*, v_{57}, v_{46}, v_{23}, q_1^*\}=
\ft18 \{\lambda_1, \lambda_1, \lambda_1, \lambda_1,
\lambda_2, \lambda_2, \lambda_2, -\lambda_2\}\ ,\nonumber\\
&&\mu=\ft12 \lambda_{12}\, \{0_4, 1_{24}, 2_4\}\ ,
\nonumber\\
&&\nonumber\\
\Delta = 2:&&m=\fft{\lambda_1 + \lambda_2}{2\sqrt2}\ ,\qquad
\{p_1, u_{12}, q_1, v_{12} \} = \ft{1}{4\sqrt{2}}
\{\lambda_1,\lambda_1, \lambda_2, \lambda_2\}\ ,\label{dyonicd4}\\
&&\mu = \sqrt2\{(\lambda_1 + \lambda_2 - \lambda_{12})_8,\,
(\lambda_1 + \lambda_2 + \lambda_{12})_8,\,
(\lambda_1 + \lambda_2)_{16}\}\ ,\nonumber
\eea
where $\lambda_{12}\equiv \sqrt{\lambda_1^2 + \lambda_2^2}$.
Note that for $\Delta =1$, the solution, is already intrinsically dyonic of
the first type even when $\lambda_1=0$ or $\lambda_2=0$, as we
discussed previously.  In fact for this solution both the metric and the
eigenvalues of the Bogomol'nyi matrix are identical to those of the dyonic
solution of the first type, even though the two Bogomol'nyi matrices are
different.   In particular, the solution always preserves $\ft14$ of the
supersymmetry, regardless of the values of $\lambda_1$ and $\lambda_2$. For
$\Delta=2$, on the other hand, we can have zero eigenvalues only for the
following three cases: $\lambda_1=0$, $\lambda_2=0$ or $\lambda_1 =
-\lambda_2$.  The first two cases correspond to the purely solitonic and
purely elementary solutions which preserve $\ft14$ of the supersymmetry. The
third case gives rise to a massless black hole (which has been discussed in
\cite{cvetic}), which preserves $\ft12$ of the supersymmetry.  However, some
of the eigenvalues are negative in this case.

\section{$p$-branes for 1-form field strengths}

     The analysis for the $p$-brane solutions with 1-form field strengths is
analogous to that for the higher-degree field strengths.  The main
difference is that one can only make a solitonic ansatz for a 1-form field
strength, since the elementary ansatz (\ref{fansatz}) is not defined when
the field strength has only one index.  The solitonic solutions are
$(D-3)$-branes.  In $D\ge 6$, the 1-forms are given by $F_{\sst{M}ijk}$ and
${\cal F}_{\sst{M}i}^{(j)}$, and the corresponding solitonic Page charges
are denoted by $v_{ijk}$ and $q_{ij}$.  In $D=5$, there is an extra 1-form
coming from the dualisation of the 4-form $F_{\sst{MNPQ}}$, whose Page
charge is denoted by $v^*$.  In $D=4$, extra 1-forms come instead from the
dualisation of the seven 3-forms $F_{\sst{MNP}i}$, and the corresponding
Page charges are denoted by $v_i^*$.   The Bogomol'nyi matrix is given by
\be
{\cal M} = m\oneone + \ft16 v_{ijk} \Gamma_{\hat1\hat2ijk} +\ft12 q_{ij}
\Gamma_{\hat1\hat2ij} + v^*\, \Gamma_{012}  + v_i^*\, \Gamma_{01i}\ ,
\label{bog1form}
\ee
where the $v^*$ and $v^*_i$ terms appear only in $D=5$ and $D=4$
respectively.

     The number of $p$-brane solutions for 1-form field strengths is far
greater than the previous cases.  We shall first present the supersymmetric
solutions.  Like the $p$-brane solutions for higher-degree field strengths,
the solutions with only one participating 1-form field strength also have
$\Delta = 4$, and they all preserve half the supersymmetry.  The first
example of such a solution is the solitonic 6-brane in $D=9$ \cite{lpss},
which is the highest dimension for 1-forms.  There are a total of 8
inequivalent supersymmetric solutions, given by
\bea
\Delta =4\ ,&& \{q_{12}\} ,\qquad \mu=2m \{0_{16}, 1_{16}\}
\ ,\nonumber\\
\Delta = 2\ ,&& \{q_{12}, v_{123}\}\ ,\qquad \mu =m
\{0_8, 1_{16}, 2_{8}\}\ ,\nonumber\\
\Delta =\ft43\ ,&& \{q_{12}, q_{45}, v_{123}\}\ ,\qquad
\mu=\ft23 m \{ 0_4, 1_{12}, 2_{12}, 3_4\}\ ,\nonumber\\
\Delta =1\ ,&& \{q_{12}, q_{45}, v_{123}, v_{345} \}\ ,
\qquad \mu =m\{ 0_4, 1_{24}, 2_4\}\ ,\nonumber\\
\Delta =1\ ,&& \{q_{12}, q_{34}, q_{56}, v_{127}\}\ ,
\qquad \mu =\ft12 m\{0_2, 1_8, 2_{12}, 3_8, 4_2\}\ ,\label{oneformsusy}\\
\Delta = \ft45\ ,&& \{ q_{12}, q_{34}, q_{56}, v_{127}, v_{347} \}\ ,
\qquad \mu= \ft25 m\{ 0_2, 1_2, 2_{12}, 3_{12}, 4_2, 5_{2}\}\ ,\nonumber\\
\Delta = \ft23\ ,&& \{q_{12}, q_{34}, q_{56}, v_{127}, v_{347}, v_{567}\}
\ ,\qquad \mu=\ft13m \{ 0_2, 2_6, 3_{16}, 4_6, 6_2\}\ ,\nonumber\\
\Delta =\ft47\ ,&& \{q_{12}, q_{34}, q_{56}, v_{127}, v_{347}, v_{567},
v^*_7\}\ ,\qquad \mu=\ft27 m \{0_2, 3_{14}, 4_{14}, 7_2\}\ .\nonumber
\eea
It is easy to verify that all these solutions satisfy the constraints implied
by both the ${\cal L}_{\sst{FFA}}$ term and the Chern-Simons modifications
to the field strengths.  The squares of the Page charges listed above are
all equal in each case.  The first case occurs in $D\le 9$, and preserves
$\ft12$ the supersymmetry; the second occurs in $D\le 8$, and preserves
$\ft14$ of the supersymmetry. The third and the fourth both occur in $D\le
6$.  They both preserve $\ft18$ of the supersymmetry, but their
non-vanishing eigenvalues are different. The last four solutions all occur
in $D=4$ only, and they all preserve $\ft1{16}$ of the supersymmetry, but
their non-vanishing eigenvalues are different.  Note that we have two cases
with $\Delta=1$, with one preserving $\ft18$ of the supersymmetry, and the
other preserving $\ft1{16}$ instead.  Some analogous solutions in the four
dimensional $N=4$ heterotic string were obtained in \cite{dfkr}.

      There are a great number of non-supersymmetric solutions.  For
example for $N=2$ participating field strengths, in addition to the
supersymmetric solution with $\Delta=2$, non-supersymmetric  solutions exist
with $\Delta = 1, 3$.  Both occur in $D\le 8$.  For $N=3$ field strengths,
$\Delta =\ft83, \ft{12}7, \ft45, \ft8{11}, \ft25$, with the first two
occurring in $D\le 7$ and the additional ones occurring only when $D\le 6$.
As the number of participating field strengths increases, which occurs as we
descend to lower dimensions, the number of non-supersymmetric
solutions increases significantly.  For example there are 43 different
$\Delta$ values for six field strengths in $D=5$.  However it is a
straightforward exercise to enumerate them by computer, using the procedure
discussed in section 3.  It involves solving eqn.\ (\ref{fequation2}) for
all possible $N\times N$ submatrices $A_{\alpha\beta}$ of the dot products
of the dilaton vectors, with $N\le (11-D)$.   We shall not present these
results in this paper.

\section{Conclusions}

     In this paper, we have looked at $p$-brane solutions in the maximal
supergravities in $11\ge D\ge 4$ dimensions.  Our starting point was to
obtain the complete bosonic Lagrangians {\it via} Kaluza-Klein dimensional
reduction from $D=11$.  Our construction of the Lagrangians distinguishes
the $(11-D)$ dilatonic scalar fields from the 0-form potentials for the
1-form field strengths.  We then consistently truncate the bosonic
Lagrangian to the form (\ref{genlag}), with one dilatonic scalar and one
$n$-form field strength.  The $p$-brane solutions involve only the metric,
the dilatonic scalar (which is a linear combination of the original $(11-D)$
dilatonic scalars), and the field strength (which is a linear combination of
the original field strengths).   We also discussed the further constraints
on the $p$-brane solutions implied by the terms coming from the dimensional
reduction of ${\cal L}_{\sst{FFA}}$, and by the Chern-Simons modifications of
the field strengths, since we focussed only on $p$-brane solutions not
involving these terms.   Classifying $p$-brane solutions of this kind
reduces to a problem in linear algebra, as we discussed in section 3.

     The metrics of these $p$-brane solutions depend only on the degree of
the field strength and the value of $\Delta$ given by (\ref{avalue}).
Different consistent truncations lead to different values of $\Delta$.
Another factor that characterises a $p$-brane solution is its supersymmetry.
The supersymmetry of a solution depends on both the value of $\Delta$ and the
configuration of the participating field strengths.  In order to study the
supersymmetry of the $p$-brane solutions, we obtained Nester forms in lower
dimensions using Kaluza-Klein dimensional reduction from $D=11$.  From the
Nester form, we obtain the Bogomol'nyi matrix, whose zero eigenvalues
correspond to the unbroken supersymmetries.  In fact a $p$-brane solution is
also characterised by the eigenvalues of its Bogomol'nyi matrix.  As we
descend through the dimensions, the number of field strengths grows rapidly,
and, as a consequence, the number of inequivalent $p$-brane solutions
proliferates.  However, most of these solutions are non-supersymmetric,
{\it i.e.}\ the corresponding Bogomol'nyi matrices have no zero eigenvalues.
Although the solutions with some unbroken supersymmetry are of primary
interest,  we have also given a fairly exhaustive classification of the
non-supersymmetric solutions too.  Specifically, we have presented all the
inequivalent $p$-brane solutions using 4-form, 3-form and 2-form field
strengths in all dimensions $D \ge 4$, and some examples of $p$-brane
solutions using 1-form field strengths.  For the supersymmetric solutions,
on the other hand, we have exhaustively studied all field strengths in all
dimensions $\ge 4$.

      The values of $\Delta$ for the supersymmetric $p$-brane solutions
can be summarised as follows:
\bigskip\bigskip

\centerline{
\begin{tabular}{|c|c|c|c|c|}\hline
&\phantom{for} 4-form\phantom{for} &\phantom{for} 3-form\phantom{for}
&\phantom{for} 2-form\phantom{for} & 1-form\\ \hline\hline
$D=11$  & $\Delta = 4 $ &        &        &       \\  \hline
$D=10$   &               & $\Delta =4$ & $\Delta=4$  &   \\ \hline
$D=9$    &    &     & $\Delta =2$ & $\Delta =4$ \\ \hline
$D=8$    &    &     &  & $\Delta =2$ \\  \hline
$D=7$    &    &     &  & \\  \hline
$D=6$    &    &  $\Delta =2$   &  & $\Delta = \ft43,1'$ \\  \hline
$D=5$    &    &     & $\Delta=\ft43$ & \\  \hline
$D=4$    &    &  &$\Delta = 1'$ & $\Delta = 1, \ft45, \ft23, \ft47$\\
\hline
\end{tabular}}

\bigskip
\centerline{Table 2: Supersymmetric $p$-brane solutions}

\bigskip

\noindent
Note that we present the the values of $\Delta$ in the dimensions where
$p$-brane solutions using the indicated field strengths first arise.
Supersymmetric solutions with the same values of $\Delta$ then occur in all
lower dimensions.  Note also that in a lower dimension, higher-degree field
strengths can be dualised to lower-degree field strengths. All these
supersymmetric solutions have $\Delta = \ft4{N}$, where $N$ is the number of
field strengths participating in the solution.   The number of different
values of $\Delta$ increases as the degree of the participating field
strength decreases.  The Page charges for all these fields are equal in each
given supersymmetric solution; denoting this value by $P$, we find that it
is related to the mass per unit volume $m$ by
$$\fft{m}{P} = N\ .\eqno(7.1)$$
The supersymmetric $p$-brane solutions with $\Delta = 4, 2$ and  $\ft43$
preserve $\ft12$, $\ft14$ and $\ft18$ of the $D=11$ supersymmetry
respectively.  The eigenvalues of their Bogomol'nyi matrices are
$2m\{0_{16}, 1_{16}\}$, $m\{0_8, 1_{16}, 2_8\}$ and $\ft23 m\{0_4, 1_{12},
2_{12}, 3_{4}\}$.  For $\Delta =1$, there are two inequivalent supersymmetry
breaking patterns, which we distinguish by writing them as $\Delta =1$ and
$\Delta = 1'$.  Both the solution in $D=4$ using 2-forms and the solution in
$D=6$ using 1-forms, {\it i.e.}\ $\Delta =1'$,  preserve $\ft18$ of the
supersymmetry, and the eigenvalues are $m\{0_4, 1_{24}, 2_{4} \}$.  On the
other hand, the solution in $D=4$ using 1-forms, {\it i.e.}\ $\Delta=1$,
has eigenvalues $\ft12 m\{0_2, 1_{8}, 2_{12}, 3_{8}, 4_{2}\}$ and hence
preserves $\ft{1}{16}$ of the supersymmetry.  The remaining three cases,
with $\Delta = \ft45, \ft23$ and $\ft47$ for 1-forms in $D=4$, all preserve
$\ft1{16}$ of the supersymmetry, and have eigenvalues equal to $\ft25
m\{0_2, 1_{2}, 2_{12}, 3_{12}, 4_{2}, 5_{2}\}$, $\ft13 m\{0_2, 2_{6},
3_{16}, 4_{6}, 6_{2}\}$ and $\ft27 m\{0_2, 3_{14}, 4_{14}, 7_{2}\}$
respectively.  It is worth remarking also that in the cases of $\Delta = 4,
2, \ft43$ and $1$, the eigenvalues of the Bogomol'nyi matrices, and hence
the supersymmetries, are independent of the signs of the Page charges.  On
the other hand, in the cases of $\Delta = 1', \ft45, \ft23$ and $\ft47$,
there are two inequivalent sets of eigenvalues depending on the signs of the
Page charges, one of which (the one we have presented) includes zero
eigenvalues, whilst the other does not.

     The above supersymmetric $p$-branes are either purely elementary or
purely solitonic, or else mixed dyonic of the first type.  The
characteristic feature of these solutions is that each participating field
strength has either elementary or solitonic (but not both) contributions.
For the 3-forms in $D=6$ and 2-forms in $D=4$, we can also construct dyonic
solutions of the second type, where the canonical field
strength has both elementary and solitonic contributions, and hence each
participating field strength has the same ratio of electric and magnetic
Page charges.  In $D=6$, the supersymmetric dyonic string preserves $\ft14$
of the supersymmetry for generic non-vanishing electric and magnetic
charges.  The mass per unit length is the sum of the electric and magnetic
charges.  When one of the Page charges is set to zero, the solution reduces
to the purely elementary or purely solitonic solution. When the two Page
charges are equal, the solution reduces to the self-dual string in $D=6$
self-dual supergravity.  On the other hand, when the two Page charges are
equal and opposite, the solution, although anti-self-dual, does not reduce to
the anti-self-dual string in anti-self-dual $D=6$ supergravity, but instead
gives rise to a massless string, however with a Bogomol'nyi matrix of
indefinite signature.  Dyonic solutions of the second type also arise in
$D=4$.  The solutions are supersymmetric for $\Delta=1$ and 2.  For
$\Delta=1$, the solution preserves $\ft14$ of the supersymmetry; for
$\Delta=2$, it gives rise to a massless black hole that preserves $\ft12$ of
the supersymmetry.

\renewcommand{\theequation}{A.\arabic{equation}}
\section*{Appendix}

    In this appendix, we present the complete bosonic Lagrangian for maximal
supergravity in $D$ dimensions, obtained by dimensional reduction from
$D=11$.  Thus starting from the Lagrangian (\ref{d11lag}), we apply the
reduction procedure described in section 2, leading to
\bea
{\cal L} &=& eR -\ft12 e\, (\del\vec\phi)^2 -\ft1{48}e\, e^{\vec a\cdot \vec
\phi}\, F_4^2 -\ft{1}{12} e\sum_i e^{\vec a_i\cdot \vec\phi}\, (F_3^{i})^2
-\ft14 e\, \sum_{i<j} e^{\vec a_{ij}\cdot \vec\phi}\, (F_2^{ij})^2
\nonumber\\
&& -\ft14e\, \sum_i e^{\vec b_i\cdot \vec\phi}\, ({\cal F}_2^i)^2
-\ft12 e\, \sum_{i<j<k} e^{\vec a_{ijk} \cdot\vec \phi}\,
(F_1^{ijk})^2 -\ft12e\, \sum_{i<j} e^{\vec b_{ij}\cdot\vec\phi}\,
({\cal F}_1^{ij})^2 + {\cal L}_{\sst{FFA}}\ ,\label{A.1}
\eea
where the dilaton vectors are defined by (\ref{dilatonvec}) and ${\cal
L}_{\sst{FFA}}$ is coming from the $\hat F_4\wedge \hat F_4 \wedge \hat
A_3$ term in the $D=11$ Lagrangian.  We shall discuss this term presently.

      First we discuss the Chern-Simons modifications to the various field
strengths appearing in (\ref{A.1}).  The Kaluza-Klein reduction of an
$n$-form potential $\hat A_n$ from $D+1$ to $D$ dimensions is given by $\hat
A_n= A_n + A_{n-1}\wedge dz$, and so the 3-form potential of $D=11$
supergravity, when reduced to $D$ dimensions, is given by
\be
\hat A_3 = A_3 + A_2^i\wedge d z^i -\ft12 A_1^{ij}\wedge dz^i\wedge dz^j
-\ft16 A_0^{ijk} dz^i \wedge dz^j\wedge dz^k\ .\label{A.2}
\ee
Taking the exterior derivative, we have
\be
\hat F_4 = \td F_4 + \td F_3^i \wedge dz^i - \ft12 \td F_2^{ij} \wedge
dz^i\wedge dz^j -\ft16 \td F_1^{ijk} \wedge dz^i \wedge dz^j \wedge dz^k
\ ,\label{A.3}
\ee
where $\td F_4 = dA_3$, $\td F_2^{ij} = d A_1^{ij}$ and
$\td F_1^{ijk} = d A_0^{ijk}$ are the unmodified field strengths.  In order
to adapt the expansion (\ref{A.3}) to the vielbein basis, we re-express
$dz^i$ in terms of $h^i = dz^i + {\cal A}_1^{i} + {\cal A}_0^{ij} dz^j$,
which is the subsequent dimensional reduction of the term $(dz^i +{\cal
A}_1^i)$ arising in the Kaluza-Klein vielbein ansatz at the $i$'th step of
the reduction process. Since the sum in the last term is only over $j >i$,
it is easy to solve iteratively for $dz^i$, giving
\bea
dz^i &=& (\delta^{i\ell} - {\cal A}_0^{i\ell} + {\cal A}_0^{ij}
{\cal A}_0^{j\ell}
-{\cal A}_0^{ij} {\cal A}_0^{jk} {\cal A}_0^{k\ell} + \cdots)
(h^{\ell} -{\cal A}_1^{\ell})\ ,\nonumber\\
&=& [(1+{\cal A}_0)^{-1}]^{i\ell} (h^{\ell} -
{\cal A}_1^{\ell})\label{A.4}\\
&\equiv& \gamma^{i\ell} (h^\ell - {\cal A}_1^\ell)\ .\nonumber
\eea
Since the 0-form potentials ${\cal A}_0^{ij}$ are defined only for $i<j$,
{\it i.e.}\ ${\cal A}_0^{ij}=0$ for $i\ge j$, it follows that the binomial
expansion of $[(1+{\cal A}_0)^{-1}]^{i\ell}$ terminates with the power
$({\cal A}_0)^{10-D}$ in $D$ dimensions. Also, we have $\gamma^{ij}=0$ for
$i>j$ and $\gamma^{ij} = 1$ for $i=j$.  Thus the 11-dimensional 4-form field
strength (\ref{A.3}) can be rewritten as
\be
\hat F_4 = F_4 + F_3^i \wedge h^i- \ft12 F_2^{ij} \wedge
h^i\wedge h^j -\ft16 F_1^{ijk} \wedge h^i \wedge h^j
\wedge h^k\ ,\label{A.5}
\ee
where the Chern-Simons modified field strengths, denoted by untilded
symbols, are given by
\bea
F_4 &=& \td F_4 - \gamma^{ij} \td F_3^i\wedge {\cal A}_1^j -\ft12
\gamma^{ik}\gamma^{j\ell} \td F_2^{ij} \wedge {\cal A}_1^k\wedge
{\cal A}_1^\ell + \ft16 \gamma^{i\ell}\gamma^{jm}\gamma^{kn}
\td F_1^{ijk}\wedge {\cal A}_1^\ell \wedge {\cal A}_1^m \wedge
{\cal A}_1^n\ ,\nonumber\\
F_3^i &=& \gamma^{ji}\td F_3^j - \gamma^{ji}\gamma^{k\ell} \td F_2^{jk}
\wedge {\cal A}_1^\ell - \ft12 \gamma^{ji}\gamma^{km}\gamma^{\ell n}
\td F_1^{jk\ell}\wedge {\cal A}_1^m \wedge {\cal A}_1^n\ ,\nonumber\\
F_2^{ij} &=& \gamma^{ki}\gamma^{\ell j} \td F_2^{k\ell} -
\gamma^{ki} \gamma^{\ell j} \gamma^{mn} \td F_1^{k\ell m}\wedge
{\cal A}_1^n\ ,\label{A.6}\\
F_1^{ijk} &=& \gamma^{\ell i} \gamma^{mj} \gamma^{nk} \td F_1^{\ell mn}
\ .\nonumber
\eea
These modified field strengths are the ones that appear in the
kinetic terms in the Lagrangian (\ref{A.1}).  Similarly, the Chern-Simons
modifications to the 2-forms and 1-forms coming from the vielbein are given by
\bea
{\cal F}_2^i &=& \td {\cal F}_2^i - \gamma^{jk} \td {\cal F}_1^{ij} \wedge
{\cal A}_1^k\ ,\nonumber\\
{\cal F}_1^{ij} &=& \gamma^{kj} \td {\cal F}_1^{ik}\ ,\label{A.7}
\eea
where again the tildes denote the unmodified field strengths; $\td{\cal
F}_2^i= d{\cal A}_1^i$, $\td{\cal F}_1^{ij}= d{\cal A}_0^{ij}$.
Note that all the Chern-Simons modifications become much simpler if the
0-form potentials ${\cal A}_0^{ij}$ from the vielbein vanish, since then
$\gamma^{ij}$ is simply $\delta^{ij}$.

     Let us now consider the dimensional reduction of the $\hat F_4\wedge
\hat F_4\wedge \hat A_3$ term in the $D=11$ Lagrangian.  Since this term is
constructed without the use of the metric or vielbein, its dimensional
reduction is written in terms of the unmodified
field strengths $\td F_4$, $\td F_3^i$, $\td F_2^{ij}$ and $\td F_1^{ijk}$.
Thus from (\ref{A.2}) and (\ref{A.3}) we find that the term ${\cal
L}_{\sst{FFA}}$ is dimensionally reduced to
\bea
D=10: &&\ft12 \td F_4\wedge \td F_4 \wedge A_2\ ,\nonumber\\
D=9: &&\Big(-\ft14 \td F_4 \wedge \td F_4 \wedge A_1^{ij}-\ft12 \td F_3^i
\wedge \td F_3^j \wedge A_3\Big)\epsilon_{ij}\ ,\nonumber\\
D=8: && \Big(-\ft1{12} \td F_4\wedge \td F_4 A_0^{ijk} -\ft16 \td F_3^i\wedge
\td F_3^j \wedge A_2^k +\ft12 \td F_3^i \wedge \td F_2^{jk} \wedge
A_3\Big) \epsilon_{ijk}\ ,\nonumber\\
D=7: && \Big(-\ft16 \td F_4\wedge \td F_3^i A_0^{jkl} +\ft16 \td F_3^{i}\wedge
\td F_3^{j} \wedge A_1^{kl} +\ft18 \td F_2^{ij}\wedge \td F_2^{kl}
\wedge A_3\Big)\epsilon_{ijkl}\ ,\label{A.8}\\
D=6: && \Big(\ft1{12} \td F_4\wedge \td F_2^{ij} A_0^{klm} +\ft1{12}
\td F_3^i\wedge \td F_3^j A_0^{klm} +\ft18 \td F_2^{ij}\wedge
\td F_2^{kl} \wedge A_2^m\Big) \epsilon_{ijklm}\ ,\nonumber\\
D=5: && \Big(\ft1{12} \td F_3^i\wedge \td F_2^{jk}  A_0^{lmn} -\ft1{48}
\td F_2^{ij}  \wedge \td F_2^{kl}\wedge A_1^{mn} -\ft1{72}
\td F_1^{ijk}\wedge \td F_1^{lmn} \wedge A_3\Big)
\epsilon_{ijklmn}\ ,\nonumber\\
D=4: && \Big(-\ft1{48} \td F_2^{ij}\wedge \td F_2^{kl} A_0^{mnp} -\ft1{72}
\td F_1^{ijk}\wedge \td F_1^{lmn} \wedge A_2^p\Big)
\epsilon_{ijklmnp}\ ,\nonumber\\
D=3:&& \ft1{144}\, \td F_1^{ijk}\wedge \td F_1^{lmn}\wedge A_1^{pq}
\epsilon_{ijklmnpq}\ ,\nonumber\\
D=2: && \ft1{1296}\, \td F_1^{ijk}\wedge \td F_1^{lmn} A_0^{pqr}
\epsilon_{ijklmnpqr}\ .\nonumber
\eea
In the $p$-brane solutions that we are considering in this paper, the
contributions that these terms give to the field equations will always be
required to be zero.  This leads to the constraints (\ref{ffaconstr})
discussed in section 2.

\section*{Acknowledgement}

     We should like to thank M.J.~Duff, J.~Rahmfeld and E.~Sezgin for useful
discussions.

\end{document}